\documentclass[twoside]{article}

\usepackage{PRIMEarxiv}

\usepackage[utf8]{inputenc} 
\usepackage[T1]{fontenc}    
\usepackage{hyperref}       
\usepackage{url}            
\usepackage{booktabs}       
\usepackage{amsfonts}       
\usepackage{nicefrac}       
\usepackage{microtype}      
\usepackage{lipsum}
\usepackage{fancyhdr}       
\usepackage{graphicx}       
\graphicspath{{media/}}     
\usepackage{amsmath} 
\usepackage{float}

\usepackage{enumitem}
\usepackage{soul}
\usepackage{multirow}
\usepackage{multicol}
\usepackage{xcolor}
\usepackage{array}

\usepackage{listings}
\usepackage[capitalise]{cleveref}
\usepackage{subcaption}
\DeclareCaptionFont{mysmall}{\fontsize{8.5pt}{10pt}\selectfont}
\usepackage[font=mysmall]{caption}
\usepackage[most]{tcolorbox}
\usepackage{cuted}
\usepackage{titlesec}

\definecolor{darkgreen}{rgb}{0,0.7,0}

\newcommand{\red}[1]{{\color{black}#1}}
\newcommand{\rr}[1]{{\color{black}#1}}
\newcommand{\rro}[1]{{\color{black}#1}}

\newcommand{\myfootnotesize}{\fontsize{7pt}{8.5pt}\selectfont}

\title{URL Extraction from Scholarly Documents: \\A Cross-Format Comparative Analysis}

\author{
  Rochana R. Obadage \\
  Old Dominion University  \\
  Norfolk, VA, USA \\
  \texttt{rochana@cs.odu.edu} \\
  \And
  Lamia Salsabil  \\
  Old Dominion University  \\
  Norfolk, VA, USA \\
  \texttt{lsals002@odu.edu} \\
  \And
  Sawood Alam  \\
  Internet Archive  \\
  San Francisco, CA, USA \\
  \texttt{sawood@archive.org} \\
  \And
  Bipasha Banarjee  \\
  Virginia Tech  \\
  Blacksburg, VA, USA \\
  \texttt{bipashabanerjee@vt.edu} \\
  \And
  William A. Ingram  \\
  Virginia Tech  \\
  Blacksburg, VA, USA \\
  \texttt{waingram@vt.edu} \\
  \And
  Edward A. Fox  \\
  Virginia Tech  \\
  Blacksburg, VA, USA \\
  \texttt{fox@vt.edu} \\
  \And
  Jian Wu \\
  Old Dominion University  \\
  Norfolk, VA, USA \\
  \texttt{j1wu@odu.edu} \\
}

\begin{document}
\maketitle

\begin{abstract}

URLs in scholarly documents link to rich external resources such as datasets, software, publications, and websites. Extracting these URLs is crucial in the data preparation stage of many downstream tasks, such as link rot analysis, web crawling, and building knowledge graphs. However, existing studies often downplay this phase, simply extracting URLs from a single format, usually text directly converted from PDFs. We present a systematic study evaluating URL extraction across six input formats (text with annotation layer, LaTeX, HTML, XML, Markdown, and PNG converted from PDF). To support the evaluation, we compiled a benchmark dataset consisting of 2,338 manually annotated URLs from 200 arXiv papers spanning a wide range of domains over a 33-year period. In addition to evaluating individual file formats, we also compared 63 composite input-format combinations. Our extensive evaluations indicate that TEXTWAL achieves the best performance among single-format inputs, while TEXTWAL+LaTeX achieves the best overall URL extraction performance. The same trend is observed for URLs linking to open-access datasets and software. To further validate these findings, we apply our format-specific URL extraction pipelines to a longitudinal random sample of 364,744 arXiv papers spanning 33 years. We observe a sharp increase in URL density after 2015, along with remarkable differences in URL extraction across file formats over time. Overall, our study highlights the importance of selecting an appropriate format for URL extraction from scholarly documents. The dataset and code are publicly available at: \textit{\url{https://github.com/lamps-lab/arxiv-url-bench}}.
\end{abstract}


\section{Introduction}
\label{sec:introduction}

Open science increasingly depends on URLs embedded in scholarly papers to direct readers to datasets, software, references, project websites, and other resources that support transparency and reproducibility\footnote{We adopt the definition of \textbf{reproducibility} from \cite{national2019reproducibility}, where a finding is deemed reproducible if \textit{consistent results are obtained using the same input data, computational steps, methods, and code, and conditions of analysis}.}. Beyond supporting reproducibility, these URLs provide access to a broad ecosystem of scholarly and web resources. Accurate and complete URL extraction is therefore a fundamental prerequisite for many downstream research tasks, including link rot analysis \cite{klein2014justkeepingtrack, url_decay, link_rot}, web crawling, scholarly knowledge graph construction, resource discovery, and large-scale studies of scholarly communication. It also serves as a critical step for identifying and cataloging open-access datasets and software (OADS) referenced in the literature.

The need for reliable URL extraction becomes particularly important when working with large scholarly repositories such as arXiv \cite{arxiv, ginsparg2011yearsagotoday}, S2ORC \cite{s2orc}, and PMC \cite{ncbi_pmc}, which collectively host tens of millions of full-text papers. At this scale, extraction errors can have substantial consequences: invalid URLs reduce precision, while missed URLs lead to incomplete analyses, resource collections, and knowledge bases. Even a small fraction of deficiencies in extraction performance can propagate to thousands or millions of missing or incorrectly identified links.

\begin{figure}[htbp]
  \centering
  \setlength{\fboxsep}{4pt}
  \setlength{\fboxrule}{0.5pt}
  \fbox{\includegraphics[width=0.65\linewidth]{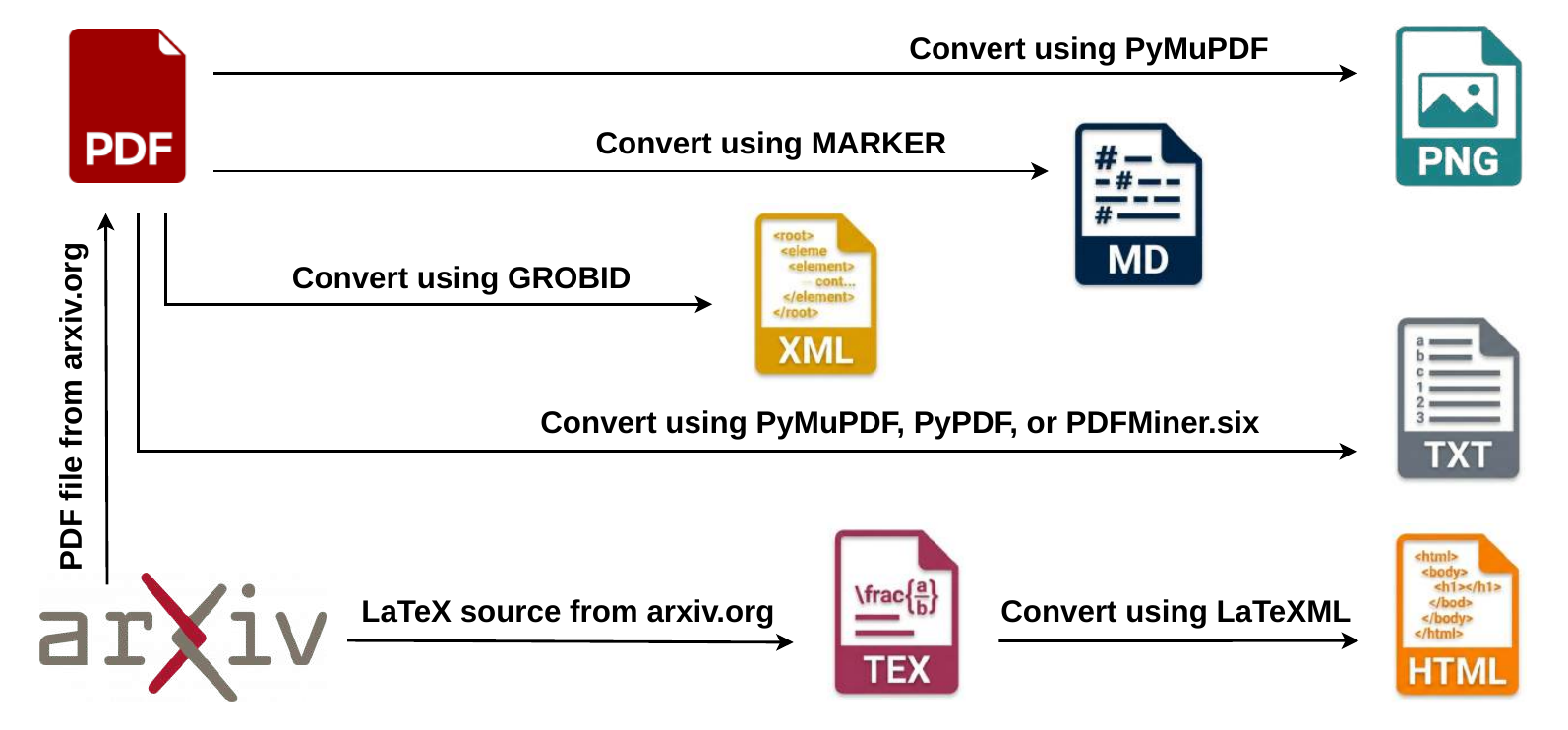}}
  \caption{Different file formats for an arXiv paper and how they are obtained for our study.}
  \label{fig:arxiv_file_formats}
\end{figure}

URL extraction from scholarly documents is commonly performed on a single document representation, typically plain text extracted from PDFs, with little consideration of alternative formats. However, as illustrated in Figure~\ref{fig:arxiv_file_formats}, URLs can be extracted from multiple representations derived from PDFs or source files (e.g., LaTeX). Because each conversion pipeline and subsequent format-specific extraction method introduces different levels of information loss, not all URLs present in the original paper may be recovered. Before scaling URL extraction to large scholarly corpora, it is therefore essential to understand how document format affects extraction quality and coverage.

To address this, we select papers from the arXiv repository, because a large fraction of papers in its repository provides multiple representations of the same paper, including PDF, LaTeX source, and HTML formats. Recently, with the growing adoption of HTML for scholarly publishing, an increasing number of arXiv papers have become available in HTML format through LaTeX-to-HTML conversion \cite{githubGitHubBrucemillerLaTeXML}.

This repository enables us to build a dataset to systematically benchmark extraction methods designed for various formats. The evaluation results will help researchers assess the potential discrepancies from the complete URL set and choose a format that balances performance and cost. This work serves as a foundational component of a broader infrastructure initiative aimed at improving the availability, discoverability, and long-term accessibility of scholarly documents.

Extracting URLs from scholarly documents presents a non-trivial technical challenge that varies substantially across document formats. The challenges arise from three aspects: document formats, extraction methods, and URL complexity. Each format introduces different complications that directly impact extraction quality. PDF text extraction suffers from systematic limitations inherent to the PDF format itself, such as line-break fragmentation introduced when text extraction tools parse multi-column layouts, which splits URLs into multiple components. These artifacts collectively degrade the performance of both text-based and image-based URL pattern matching.

LaTeX source files preserve authorial intent through explicit markup, yet extraction requires handling of macros and packages like \texttt{hyperref} that encode URLs through semantic commands. HTML and XML representations offer richer semantic structure but depend entirely on conversion pipelines, whose success varies with source quality and pipeline maturity. Markdown representations generated by neural converters offer an alternative that combines layout reconstruction with structured hyperlink preservation, but does not guarantee to preserve all URLs.

\rr{Extracting directly from PDFs also remains challenging. Vision-language models (VLMs) and large language models (LLMs) operate on page images or raw PDF bytestreams, providing a unified extraction approach. However, proprietary models are more expensive than open-weight alternatives, may hallucinate plausible but false URLs, and often lack reproducibility due to model version changes.}

Moreover, URL complexity interacts with format-specific artifacts in ways that prior literature has not systematically characterized.

Despite growing interest in extracting resource links from scholarly documents, two key limitations remain.
First, existing studies lack a large-scale, statistically robust benchmark for comparing URL extraction performance across multiple document formats.
Second, there is no formal taxonomy that characterizes URL complexity in scholarly documents and supports systematic diagnosis of extraction failures.

To address these gaps, \rro{we extend our previous work \cite{obadage2025robusturlextractionopen} and} conduct a multi-format benchmark study on 2,338 URLs manually extracted from 200 arXiv papers, stratified by publication year and document category. We propose a three-dimensional taxonomy of URL complexity and evaluate a diverse set of extraction methods, including traditional tools, open-weight VLMs, and proprietary LLMs. Our study further incorporates a cost-performance analysis and provides practical guidance for large-scale deployment.

Our main contributions are as follows:
{
\begin{enumerate}[leftmargin=*, label=\textbf{C\arabic*.}]

\item \textbf{A benchmark corpus.}
We construct a benchmark dataset of 2,338 URL instances manually identified from 200 arXiv papers, including approximately 20\% OADS-related links.
The corpus spans publication years 1992--2024 and includes diverse document types such as single-column, double-column, ETDs, and scanned papers.

\item \textbf{A three-dimensional URL complexity taxonomy.} We propose a taxonomy that characterizes URL extraction difficulty along three orthogonal dimensions: lexicographical, spatial, and representational complexity.

\item \textbf{Extensive evaluation.}
We evaluate six individual formats and 63 union-format combinations, together with four frontier VLM/LLM-based extraction methods (Qwen2-VL-7B, MiniCPM-o-2.6, DeepSeek-VL-7B, and Claude Haiku 4.5), applied on selected formats.

\item \textbf{A practical decision framework.}
We derive actionable recommendations for selecting formats and extraction methods under scalability, cost, and recall requirements.

\item \textbf{Large-scale longitudinal analysis.}
We apply our extraction pipeline to 467,767 arXiv papers across 33~years (1992--2024) and analyze temporal trends in URL usage. We release a dataset consisting of $\sim$360k arXiv papers each available in all six formats, enabling reproducible, format-comparable longitudinal comparison.

\end{enumerate}
}

\section{Related Work}
\label{sec:related}

\paragraph{\textbf{\small URL and Reference Extraction from Scholarly Documents.}}
URL extraction is a fundamental task to digital libraries and bibliometrics. Early approaches rely on regex-based extraction from text, as used at scale in S2ORC \cite{s2orc}, offering interpretability but limited robustness. Layout-aware systems such as GROBID \cite{lopez2009grobid} improve precision via document structure modeling, while PDF-to-text tools (PyMuPDF \cite{PyMuPDF}, PyPDF \cite{PyPDF}, PDFMiner.six \cite{pdfminer.six}) reflect different trade-offs in speed, accuracy, and layout preservation. Prior work typically evaluates tools in isolation, leaving little guidance on format selection or combinations, which this work addresses. 

\paragraph{\textbf{\small Open-Access Datasets and Software Discovery.}}
The OADS taxonomy captures dataset and software references critical for reproducibility \cite{salsabil2025context}. Salsabil~et~al. \cite{salsabil2025context} define six OADS categories, \rr{while systems such as DataStet \cite{datastet} identify explicit and implicit dataset mentions in scientific documents.} Escamilla~et~al. \cite{escamilla2023itsjustgithubidentifying} show that OADS URLs have a distinctive structure enabling specialized detection. Ajayi~et~al. \cite{kenny-icdar-2023} further highlight accessibility issues of OADS URLs in AI reproducibility \cite{acm-rep-24, 11363696}. However, no prior work systematically studies how document format choice affects OADS extraction, which we address via format-stratified evaluation.

\vspace{-1mm}
\paragraph{\textbf{\small Vision Language Models and Large Language Models.}}
\rr{Open-weight VLMs (Qwen2-VL \cite{placeholder2024qwen}, MiniCPM \cite{placeholder2024minicpm}, DeepSeek-VL \cite{placeholder2024deepseek}) operate directly on rendered pages, avoiding text extraction artifacts, while proprietary LLMs (e.g., GPT-5, Claude) extract information directly from PDFs but introduce cost, hallucination, and reproducibility concerns. A recent work, MOLE \cite{mole-2025}, uses schema-guided prompting to extract structured metadata from scientific papers. However, no prior work systematically compares VLMs and LLMs against structure-based pipelines for URL extraction. 
}

\vspace{-1mm}
\paragraph{\textbf{\small Link Rot and Temporal Analysis of Scholarly URLs.}}
Link rot threatens reproducibility as URLs disappear due to deletion or restructuring. Prior studies have documented substantial decay in scientific and software-related URLs \cite{klein2014justkeepingtrack, Hennessey2013, david2025github}. However, previous works have not performed longitudinal and comparative analyses of URLs extracted across different formats.

\section{Dataset Construction}
\label{sec:dataset}


We selected arXiv (snapshot: January 2024) as our source corpus and derived formats from either PDFs or LaTeX source files ({\small Figure \ref{fig:arxiv_file_formats}}).

\subsection{Stratified Sample Selection}
\label{sec:dataset:sampling}

We selected a 200-paper benchmark\footnote{\url{https://huggingface.co/datasets/rochanaro/hf-arxiv-url-bench}} (Table \ref{tab:sample_composition}) in which each paper is available in all six file formats (TEXT, Markdown, LaTeX, HTML, XML, and IMAGE) considered in our study. We used stratified sampling to ensure balanced temporal coverage, diverse document layouts, and representation of edge-case formats. We conducted the sampling in two phases that covered complementary time periods. For both phases, we imposed a constraint that papers must have at least three embedded URLs.

\renewcommand{\arraystretch}{1.25}
\begin{table}[h]
\centering
\small
\caption{Stratified dataset composition by document structure and time period (200 papers across two phases, with 2,420 total URLs and 2,338 unique URLs), where each paper includes all six formats: TEXT, LaTeX, XML, HTML, Markdown, and PNG.}
\label{tab:sample_composition}
\begin{tabular}{lrrr}
\toprule
\textbf{Category} & \textbf{Phase A (2016--24)} & \textbf{Phase B (1992--15)} & \textbf{Subtotal} \\
\midrule
Single-column & 45 & 40 & 85 \\
Double-column & 46 & 40 & 86 \\
ETD           &  9 & 10 & 19 \\
Scanned       &  0 & 10 & 10 \\
\midrule
\textbf{Total papers} & \textbf{100} & \textbf{100} & \textbf{200} \\
\bottomrule
Total URLs & 1,709 & 711 & 2,420 \\
\textbf{Unique URLs} & \textbf{1,667} & \textbf{671} & \textbf{2,338} \\
\bottomrule
\end{tabular}
\end{table}

\vspace{-3mm}
\subsubsection{Phase A (2016--2024):}
We stratified the initial sample of 2,250 PDF papers by year and document type, selecting 100 single-column papers, 100 double-column papers, and 50 ETDs for each year. We then converted each PDF to plain text and applied regular expression-based URL extraction, identifying 607 papers that exceeded our URL-count threshold. We then manually verified document layouts and checked format availability. After confirming the availability of LaTeX source files and successful conversion to TEXT, HTML, XML, and Markdown formats, we retained 182 papers. We randomly selected the final Phase A sample of 100 papers, consisting of 45 single-column papers, 46 double-column papers, and 9 ETDs.

\subsubsection{Phase B (1992--2015):}
Because of the relatively low number of ETDs and no scanned documents selected in Phase A, we deliberately selected more documents for each category to account for 
scanned PDFs and submissions without LaTeX allowed into arXiv in its early years. We stratified the initial sample of 3,840 PDF papers by year and document type, selecting 50 single-column papers, 50 double-column papers, 30 ETDs, and 30 scanned papers for each year.
After applying the same verification procedure as in Phase A, we retained 285 papers. From this set, we randomly selected 100 papers. Table \ref{tab:sample_composition} shows the distribution across document categories.

In our paper sample, we have papers from Computer Science (65); Physics (105); Mathematics (19); Electrical Engineering and Systems Science (5); and Statistics (4); with smaller representations in Quantitative Finance (1) and Quantitative Biology (1).

\subsection{File Format Conversion and URL Extraction}
\label{sec:dataset:formats}

For each sampled paper, we obtained multiple document formats (Figure~\ref{fig:arxiv_file_formats}) and applied format-specific URL extraction approaches (Table~\ref{tab:format_tools}). We collected native PDF and LaTeX sources from arXiv and derived other formats, including plain text (TEXT), text with annotation layers (TEXTWAL), structured markup formats (HTML, XML, and Markdown), and PNG images (IMAGE). The details of the document formats and URL extraction methods follow.

\renewcommand{\arraystretch}{1.25}
\begin{table}[h]
\small
\centering
\caption{Document formats, conversion tools, and URL extraction approaches used in the benchmark pipeline.}
\label{tab:format_tools}
\begin{tabular}{lll}
\toprule
\textbf{Format} & \textbf{Conversion Tool} & \textbf{URL Extraction Approach} \\
\midrule
PDF                          & ---           & Manual annotation (Ground Truth) \\
TEXT                         & PyMuPDF (v1.26.4)               & Rule-based RegEx \\
TEXTWAL      & PyMuPDF (v1.26.4)               & Rule-based RegEx \\
TEXTWAL     & PyPDF (v6.6.2)                  & Rule-based RegEx \\
TEXTWAL      & PDFMiner.six ({\footnotesize v260107})        & Rule-based RegEx \\
TEXTWAL    & PyMuPDF (v1.26.4)     & LLM \texttt{\footnotesize (Claude Haiku 4.5)} \\
XML                          & GROBID (v0.8.1)  & Attribute search \texttt{\footnotesize (``target'' tag)} \\
Markdown                     & Marker (v1.10.2)   & Rule-based RegEx \\
IMAGE     & PyMuPDF (v1.26.4)    & VLM \texttt{\footnotesize (Qwen2-VL-7B-Instruct)} \\
IMAGE    & PyMuPDF (v1.26.4)     & VLM \texttt{\footnotesize (MiniCPM-0 2.6)}\\
IMAGE    & PyMuPDF (v1.26.4)     & VLM \texttt{\footnotesize (DeepSeek-VL-7B-Chat)}  \\
LaTeX                        & ---                              & Attribute search (\texttt{\footnotesize \textbackslash url*}) + RegEx \\
HTML                         & LaTeXML (v0.8.8)                & Attribute search \texttt{\footnotesize (<a href="">)} \\
\bottomrule
\end{tabular}
\end{table}

\subsubsection{Document Formats}
\label{sec:approaches:formats}

\paragraph{\textbf{TEXT}}

We employed three widely used PDF-to-text extraction tools: PyMuPDF \cite{PyMuPDF}, PyPDF \cite{PyPDF}, and PDFMiner.six \cite{pdfminer.six}. We initially applied URL extraction to the visible page text produced by each converter and treated the output of each tool as a variant of the TEXT format. These tools differ in extraction speed, layout preservation, and robustness to different document structures. Our preliminary analysis showed that plain-text (TEXT) extraction alone missed a substantial number of valid URLs because many hyperlinks are stored in PDF annotation layers rather than in the visible text. 

To address this limitation, we extended each text representation with additional PDF structures, creating the \textbf{TEXTWAL} (Text With Annotation Layer) format. TEXTWAL combines page text with hyperlink annotations, optional content groups, document metadata, PDF object dictionaries, and raw PDF byte streams. URLs extracted from all layers are merged and deduplicated prior to evaluation.
\vspace{-1.5mm}

\paragraph{\textbf{LaTeX}}

We retrieved LaTeX source files from arXiv source archives and parsed URLs from both \texttt{.tex} and bibliography files (\texttt{.bbl}). LaTeX provides direct access to the author-created content and preserves hyperlink commands during manuscript preparation.
\vspace{-1.5mm}

\paragraph{\textbf{HTML}}

We converted LaTeX sources to HTML using LaTeXML \cite{githubGitHubBrucemillerLaTeXML}, the same tool used in arXiv’s rendering pipeline. HTML provides a structured representation in which hyperlinks are often explicitly encoded as document elements and attributes.
\vspace{-1.5mm}

\paragraph{\textbf{XML}}

We converted PDFs to TEI-XML using GROBID \cite{GROBID} in full-text mode. We selected GROBID because of its strong performance in scholarly PDF parsing and its rich output as structured TEI representations of scientific documents.
\vspace{-1.5mm}

\paragraph{\textbf{Markdown}}

We converted PDFs to Markdown using Marker \cite{Datalabtomarker}, a neural PDF-to-Markdown conversion system. Marker was chosen for its strong ability to preserve document structure (e.g., headings, lists, and links) while producing clean, readable Markdown suitable for downstream parsing.
\vspace{-1.5mm}

\paragraph{\textbf{IMAGE}}

We converted PDF pages to PNG images (IMAGE) using PyMuPDF with a 2$\times$ rendering zoom. This representation preserves the visual appearance of the original document and serves as input to vision-language models for image-based URL extraction.

\subsubsection{URL Extraction Methods}
\label{sec:approaches:methods}

\paragraph{\textbf{Rule-Based RegEx Extraction}}

We used a consistent RegEx-based extraction engine for all applicable file formats. The pattern follows RFC 3986\footnote{\url{https://datatracker.ietf.org/doc/html/rfc3986}} conventions while accommodating formatting artifacts commonly found in scholarly documents. It supports both standard URLs and frequently occurring scheme-less academic references (e.g., \texttt{doi.org/...}), handles surrounding delimiters, and removes trailing punctuation introduced by prose. The complete pattern is provided in Appendix ~\ref{app:regex}
.

\paragraph{\textbf{Attribute-Based Extraction}}

For structured formats, we supplemented RegEx matching with format-specific attribute and tag parsing. In LaTeX, we extracted URLs from hyperlink commands such as \texttt{\textbackslash url{}}, \texttt{\textbackslash urladdr{}}, and \texttt{\textbackslash href{}{}}. In HTML, URL extraction primarily relies on \texttt{href} attributes within anchor elements (\texttt{<a href=``...''>}), supplemented by a secondary RegEx pass over the rendered text. In XML, URLs are extracted from \texttt{target} attributes within \texttt{<ref target=``...''/>} elements. For Markdown, extraction combines explicit parsing of Markdown hyperlink syntax (\texttt{[text](url)}) with a secondary RegEx pass over the document body. 

\paragraph{\textbf{Vision-Language Model Extraction}}
\label{sec:approaches:vlm-open}

We evaluated three open-weight vision-language models (VLMs): Qwen2-VL-7B-Instruct, MiniCPM-o-2.6, and DeepSeek-VL-7B, representing strong small-to-mid-scale models with low computational requirements and competitive performance on image-based benchmarks. Each page image was processed independently using a standardized prompt (Appendix ~\ref{sec:app:vlm-prompt}) 
to extract all URLs across headers, footers, tables, references, and fragmented text spans. We used deterministic decoding (temperature = 0) for all experiments.

\paragraph{\textbf{Large Language Model Extraction}}
\label{sec:approaches:llm-proprietary}

We evaluated Claude Haiku 4.5 (\texttt{v20251001}) using the PyMuPDF TEXTWAL representation as input. The prompt instructed extraction of all URLs, reconstruction of fragmented URLs, and normalization of DOI references, returning a deduplicated URL list (Appendix ~\ref{sec:app:llm-prompt}).

\subsection{URL Complexity Taxonomy}
\label{sec:dataset:complexity}

To support fine-grained diagnosis of complex URL extraction failures, we introduce a three-dimensional taxonomy (Appendix ~\ref{app:url_complexities_taxonomy}).
\subsubsection{Lexicographical Complexity}
\label{sec:dataset:complexity:lex}

Lexicographical complexity captures structural and character-level variation in URL composition. URLs may incorporate various protocol schemes including \texttt{http}, \texttt{https}, \texttt{ftp}, \texttt{sftp}, \texttt{git}, and \texttt{magnet}. URLs may contain subdomains (e.g., \texttt{docs.example.com}), Unicode characters, complex paths with multiple directory levels, port numbers, query strings and URL fragments, percent-encoding sequences for special characters, and trailing slashes. Conversely, certain URLs omit protocols entirely (scheme-less URLs). 
\subsubsection{Spatial Complexity}
\label{sec:dataset:complexity:spatial}

Spatial complexity captures how document layout and placement affect URL extraction. URLs may appear as a continuous sequence of characters on a single line, be split across multiple lines, span page boundaries, or be located within footnotes, captions, or reference structures.
\subsubsection{Representational Complexity}
\label{sec:dataset:complexity:repr}

Representational complexity captures how URLs are encoded or concealed within documents. Certain URLs appear in visible text. Others appear only in the PDF annotation layer as hyperlinks. In certain cases, URLs are present only in document metadata fields, and in others they are shortened using URL-shortening services, requiring de-referencing to identify their original targets.

\subsection{Manual URL Annotation and Ground Truth}
\label{sec:dataset:annotation}

We conducted a comprehensive manual annotation of all URLs in the corpus we curated. \rro{Two annotators independently reviewed paper PDFs (100 each) to extract URLs (referred to as \textbf{valid URLs})}. 
For each URL, the annotators recorded the paper identifier, page number, complete URL string, section category (we assigned each URL to one of 11 section categories: Abstract, Introduction, Related Works \& Background, Methodology, Results \& Discussion, Conclusions, Future Work, References, Header, Footer, and Title \& Authors), complexity labels across three dimensions, and the OADS category following the provided guidelines\footnote{\url{https://github.com/lamps-lab/arxiv-url-bench/blob/main/documents/arxiv_url_annotation_guidelines.pdf}}.

\rro{To ensure consistency, all annotations underwent an internal review process, where a subset of papers (50\%) was independently re-checked by a second annotator. The two annotators' independent labels agreed on 99.2\% of URLs prior to resolution, and the remaining disagreements were resolved through discussion. In addition, we performed targeted spot checks across different sections to reduce systematic omissions. Our final ground truth contains all URL instances from which we extracted and verified 2,338 unique URLs.}

\subsection{Sample Characteristics}
\label{sec:dataset:distribution}

\begin{figure}[ht]
  \centering
  \fbox{\includegraphics[width=0.80\linewidth]{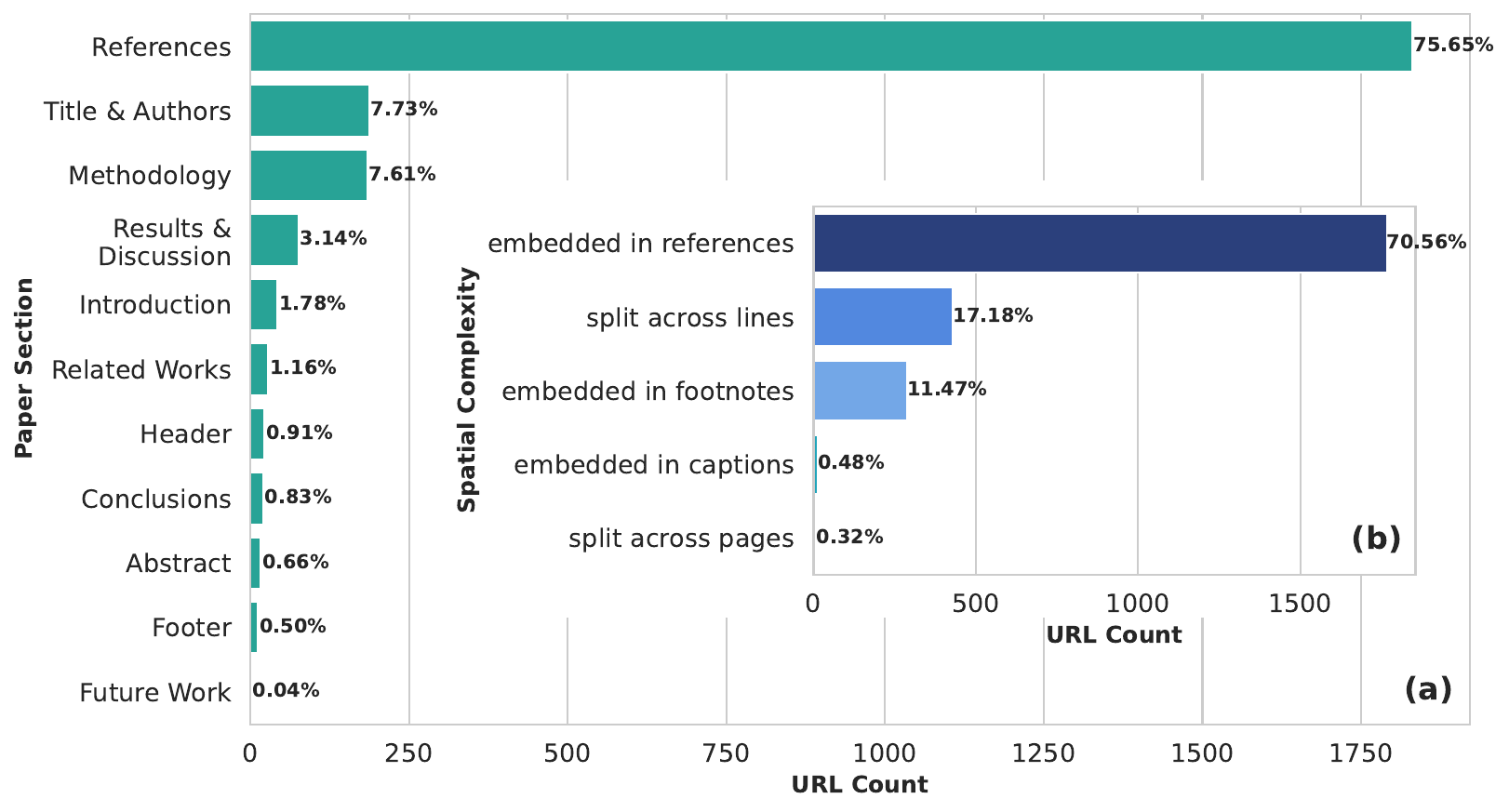}}
  \caption{The distribution of URLs across different section categories and spatial complexities. (a): The breakdown of URLs by section labels. (b) (the inset): The frequency of specific Spatial Complexity types.}
  \label{fig:spatial_complexities}
\end{figure}

Our benchmark\footnote{\url{https://huggingface.co/datasets/rochanaro/hf-arxiv-url-bench/blob/main/arxiv-200-benchmark/arxiv_urls_1992_2024_gt.csv}} comprises 2,420 URL instances in total, including 2,338 unique URLs (Table \ref{tab:sample_composition}). As shown in Figure \ref{fig:spatial_complexities}(a), URLs are distributed across all sections and locations, with the majority appearing in the References section. The others appear in Title \& Authors (7.73\%), Methodology (7.61\%), and Results \& Discussion (3.14\%) 
sections.

Our analysis of spatial complexity (Figure \ref{fig:spatial_complexities}(b)) further reveals that most URLs are embedded within reference entries, while line-split URLs account for 430 instances ($\sim$18\%). Page-spanning URLs are rare (0.3\%). The URLs appearing in footnotes and figure captions account for 11.8\% and 0.4\% of the instances, respectively.

\subsection{OADS URL Categorization}
\label{sec:dataset:oads}
To further study how format impacts OADS URLs, we manually annotated all URLs into six categories (Appendix \ref{app:oads_taxonomy}). 
We then performed a post hoc evaluation of format-specific OADS URL extraction. Our corpus contains 488 OADS URLs ($\sim$21\% of all unique URLs) (Table \ref{tab:oads_distribution} 
in the Appendix).


\section{Evaluation}
\label{sec:evaluation}

\subsection{Evaluation Metrics}
\label{sec:evaluation:framework}

We adopt standard classification evaluation metrics including precision, recall, and F1 score tailored to URL extraction. The definitions of precision and recall are shown in Eq. (1) and (2). \rro{Using strict matching}, an extracted URL is considered \textit{\textbf{valid}} if it appears in the PDF and is correctly identified during manual annotation. All other outputs are considered \textit{\textbf{invalid}}, including hallucinated strings, malformed URLs, or any URL not present in the PDF.
{\small
\begin{align} 
  \text{Precision} &= \frac{\text{Correctly Extracted Valid URLs}}{\text{All Extracted URL Strings}} \\[4pt]
  \text{Recall}    &= \frac{\text{Correctly Extracted Valid URLs}}{\text{Total Valid URLs (Ground Truth)}}
\end{align}
}

In addition, we evaluate OADS URLs separately, focusing on the 488 URLs linking to open-access datasets and software in the ground truth. Since OADS URLs are identified only after extraction through matching against the annotated ground truth, we cannot reliably attribute individual extracted URL strings to the OADS subset at extraction time. While we observe the total number of extracted URL strings, we do not have a separable breakdown of true positives and false positives for URLs classified as OADS (because we did not classify them using a model). We therefore report only Recall for OADS URLs, defined as the fraction of ground truth OADS URLs successfully retrieved by each method.

\subsection{Format Combinations}
\label{sec:evaluation:combinations}

While we report performance for all 12 individual format variants, our combination analysis is based on six core file formats: TEXTWAL, LaTeX, HTML, XML, Markdown, and IMAGE (Table \ref{tab:combinations} with full results provided in Appendix \ref{app:all_format_combinations}). 
We reduce the effective number of format variants by retaining only the best-performing representative from each format family. For example, we select TEXTWAL extracted by PyMuPDF as the representative of the TEXTWAL family because it achieves the highest URL extraction performance (Section \ref{sec:results:text-tools}) among the other methods that extract urls from TEXTWAL. For open-weight VLMs, we include the best-performing model, Qwen2-VL-7B. This consolidation reduces the format combination space from $2^{12}$ to $2^{6}$ while preserving diversity across fundamentally distinct file formats and extraction paradigms. For each combination, we use a union-based aggregation of extracted URLs with deduplication. 

\subsection{Statistical Testing}
\label{sec:evaluation:statistics}

\subsubsection{Bootstrap Confidence Intervals}
We use the individual document (identified by \texttt{arxiv\_id}) as the unit of analysis. We estimate 95\% confidence intervals for Precision, Recall, and F1 using paired bootstrap resampling at the document level. For each of 10,000 bootstrap iterations \cite{Pattengale2010-du}, we sample 200 documents with replacement and recompute the evaluation metrics. We report results as \textit{$\text{metric}~[\text{lower},\,\text{upper}]$} in Table \ref{tab:combinations}.

\subsubsection{Friedman Test for Global Format Comparison}

To assess whe\-ther statistically significant performance differences exist among the six individual formats, we apply a Friedman test using per-paper Recall scores, treating format as the within-subject factor. The test is applied independently to both All-URL Recall and OADS Recall (see Appendix \ref{app:stats}). 

\subsubsection{Wilcoxon Signed-Rank Tests for Pairwise Comparisons}

Following a significant Friedman result, we conduct pairwise Wilcoxon signed-rank tests on per-paper \emph{Recall} scores, comparing each of the 62 remaining format combinations with the best-performing single-format baseline for recall. Holm-corrected $p$-values for all comparisons are reported in Table \ref{tab:combinations} with more details shown in Appendix \ref{app:all_format_combinations}.


\section{Results}
\label{sec:results}

\subsection{PDF to Text Extractor Comparison}
\label{sec:results:text-tools}

We first compare three tools that automatically extract text from PDF documents to identify the best tool based on the number of URLs that can be extracted using the same regular expressions. The results in Table \ref{tab:text_tool_comparison} reveal clear trade-offs between extraction quality and computational efficiency. PyMuPDF achieves a recall of 0.81 and the highest precision of 0.59, extracting URLs at 0.19 s/PDF. The TEXT output format misses most URLs (558 vs. 1,887 for TEXTWAL), demonstrating the necessity of incorporating URLs from annotation layers. PyPDF attains a slightly higher recall (0.83) but lower precision (0.58) and is about 10$\times$ slower (1.99 s/PDF). PDFMiner.six matches PyMuPDF in recall and precision (0.81 and 0.59) but is much slower (4.37 s/PDF, $\sim$23$\times$ slower) without any performance gains. Overall, PyMuPDF offers the best balance of accuracy and speed and is therefore preferred for TEXTWAL.

\begin{table}[h]
\centering
\small
\caption{URL extraction performance comparison of PDF text extraction tools. P = precision; R = recall; F1 = F1-score. TEXT = text-only extraction; TEXTWAL = text + annotation layer extraction.}
\label{tab:text_tool_comparison}
\begin{tabular}{lccccc}
\toprule
\textbf{Tool} & \textbf{Output Format} & \textbf{P} & \textbf{R} & \textbf{F1} & \textbf{Avg.\ (s/PDF)} \\
\midrule
PyMuPDF      & TEXT     & 0.53 & 0.24 & 0.33 & 0.13s \\
\textbf{PyMuPDF}      & TEXTWAL  & 0.59 & 0.81 & 0.68 & \textbf{0.19s} \\
PyPDF        & TEXTWAL  & 0.58 & 0.83 & 0.68 & 1.99s \\
PDFMiner.six & TEXTWAL  & 0.59 & 0.81 & 0.68 & 4.37s \\
\bottomrule
\end{tabular}
\end{table}

\subsection{LLM and VLM Comparison for URL Extraction}
\label{sec:results:llm-vlm}
We compare three open-weight VLMs and a commercial LLM. Although open-weight VLMs perform strongly on object recognition benchmarks \cite{STATUS-Bench}, they do not perform equally well on URL extraction (Table \ref{tab:llm_vlm_comparison}), with Qwen2-VL-7B achieving the best precision and recall. Claude Haiku 4.5 achieves a precision of 0.53 and a recall of 0.38 at an estimated cost of \$0.0285 per paper. Therefore, despite lower hallucination rates than the open-weight models, Claude Haiku 4.5 still underperforms the rule-based TEXTWAL pipeline (Tables \ref{tab:text_tool_comparison}, \ref{tab:llm_vlm_comparison}).

\begin{table}[h]
\centering
\small
\caption{URL extraction performance comparison across 4 LLM/VLMs. 
P = precision; R = recall; F1 = F1-score; Hal.\ R\% = hallucination rate, computed as 
$(N_{\text{extracted URLs not in PDF}} / N_{\text{extracted URLs}}) \times 100\%$.}

\label{tab:llm_vlm_comparison}
\begin{tabular}{lcccccc}
\toprule
\textbf{Model} & \textbf{Format} & \textbf{P} & \textbf{R} & \textbf{F1} & \textbf{Hal.\ R\%} \\
\midrule
Qwen2-VL-7B       & IMAGE   & 0.11 & 0.34 & 0.17 & 89.35\% \\
MiniCPM-o-2\_6    & IMAGE   & 0.02 & 0.21 & 0.03 & 98.34\% \\
DeepSeek-VL-7B    & IMAGE   & 0.00 & 0.00 & 0.00 & 99.99\% \\
Claude Haiku 4.5  & TEXTWAL & 0.53 & 0.38 & 0.44 & 47.19\% \\
\bottomrule
\end{tabular}
\end{table}

Overall, neither VLM nor LLM approaches outperform the rule-based TEXTWAL pipeline. With superior recall, deterministic behavior, low cost, and scalability, TEXTWAL remains the preferred format for large-scale scientific PDF URL extraction.


\subsection{File Format Extraction Performance}
\label{sec:results:format-performance}

In this section, we compare the URL extraction performance across six individual formats---TEXTWAL, Markdown, LaTeX, HTML, XML, and IMAGE---as well as their 63 possible format combinations. The results are shown in Table \ref{tab:combinations}. More detailed results are provided in Appendix \ref{app:all_format_combinations}.

\subsubsection{Single-Format Performance}
\label{sec:results:single-format}

The formats exhibit distinct performance characteristics reflecting their underlying structures and extraction mechanisms. TEXTWAL achieves the highest single-format F1 score ($0.69$) and recall ($0.81$), recovering nearly three times as many valid URLs as plain TEXT extraction. \rr{Across all URLs, LaTeX achieves the highest precision among text-based representations through explicit hyperlink markup, while HTML and XML provide complementary coverage but lower overall performance due to conversion limitations. IMAGE performs the worst (F1 $=0.17$) because of its high false-positive rate. Overall, the formats ranked by URL extraction performance (F1 score) are:}
\texttt{\[
\text{TEXTWAL} > \text{Markdown} > \text{LaTeX} > \text{HTML} \approx \text{XML} > \text{IMAGE}
\]}

\subsubsection{Multi-Format Combination Analysis}
\label{sec:results:combinations}

Combining formats thr\-ough URL set union consistently increases recall at the expense of precision. Across nearly all high-performing combinations, TEXT\allowbreak WAL serves as the essential core format.

Among two-format combinations, TEXTWAL+LaTeX provides the best balance between precision and recall, increasing recall from $0.81$ to $0.88$ without reducing F1. TEXTWAL+Markdown achieves the highest two-format recall ($0.90$). Format combinations without TEXTWAL perform substantially worse, confirming the importance of annotation-layer information.

For three-format combinations, TEXTWAL+LaTeX+XML achieves the highest overall F1 ($0.68$), while TEXTWAL+LaTeX+Markdown provides slightly higher recall. The combination LaTeX+HTML +XML from our previous pilot study \cite{obadage2025robusturlextractionopen} remains weaker than any TEXTWAL-based combinations, indicating that structured formats alone cannot fully cover URLs in the annotation layer.

\begin{figure*}[h]
  \centering
  \setlength{\fboxsep}{0.5pt}\setlength{\fboxrule}{0.0pt}
  \fbox{\includegraphics[trim=1 15 1 1, clip,
    width=1.0\linewidth]{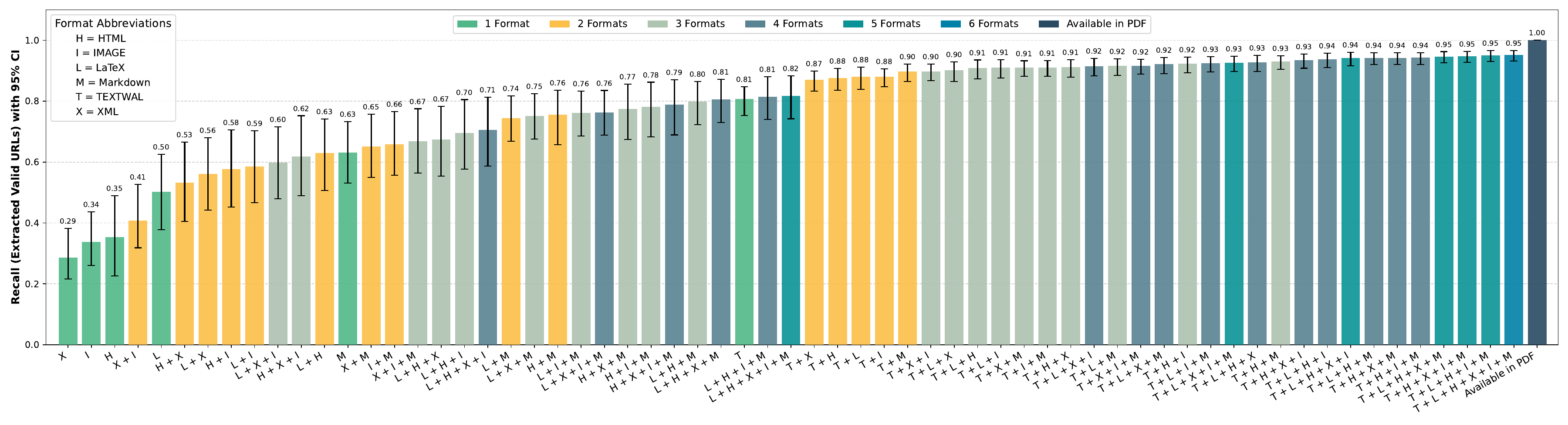}}
    \caption{\footnotesize URL extraction performance across all 63 format combinations for all valid URLs, ordered by recall. The X-axis represents the format combinations, where single-format scenarios are labeled by their full name (e.g., XML, HTML, IMAGE for Qwen2-VL), and multi-format combinations are abbreviated. 
  The "Available in PDF" category serves as the ground truth. Error bars represent 95\% confidence intervals.}

  \label{fig:url_recall_all_63}
\end{figure*}

Unioning more formats continues to improve recall but yields diminishing F1. TEXTWAL+LaTeX+HTML+Markdown provides the strongest four-format recall (Figure \ref{fig:url_recall_all_63}), but larger unions decrease precision. Incorporating IMAGE into any combination substantially reduces precision due to hallucinated URLs.
The full six-format union achieves the highest recall but the lowest F1, illustrating the fundamental precision--recall trade-off of multi-source aggregation.

In summary, TEXTWAL+LaTeX represents the best choice for URL extraction, exhibiting a substantial recall improvement over TEXTWAL alone without sacrificing F1. When additional recall is desired at lower computational cost than full ensembles, TEXTWAL+ LaTeX+HTML+XML+Markdown offers a strong five-format alternative. 

\red{Consistent with the overall results, OADS-focused retrieval benefits from multi-format combinations, with several combinations exceeding 90\% recall (Appendix: Figure ~\ref{fig:oads_recall_63}). 
TEXTWAL+LaTeX+ Markdown achieves 95\% OADS recall, while larger combinations reach up to 98\% recall by incorporating additional formats. These results suggest that broader format aggregation is beneficial when maximizing OADS retrieval coverage is the primary objective.}

\renewcommand{\arraystretch}{1.25}
\begin{table*}[ht]
{\myfootnotesize
\centering
\caption{URL extraction performance across representative format combinations.
  \#V = valid URLs extracted (from 2338); \#VO = valid OADS URLs extracted (from 488); We show the 95\% confidence intervals of precision (P), recall (R), and F1-scores. The best F1 for each group is highlighted in \textbf{bold}. \dag~= previous-study \cite{obadage2025robusturlextractionopen} comparison (LaTeX+HTML+XML);
  Sig.\ vs.\ best single-format (Wilcoxon, Holm--Bonferroni corrected):
  $^{***}p<0.001$, $^{**}p<0.01$, $^{*}p<0.05$, \textit{ns=not significant}. The table reporting results of all 63 format combinations is in Appendix  \ref{app:all_format_combinations}
  .}
\label{tab:combinations}
\begin{tabular}{p{0.15\linewidth}llllrr|llrr}
\toprule
\multirow{2}{=}{\textbf{Format Combination} \newline \textbf{(top-k by R)}}
  & \multicolumn{6}{c|}{\textbf{Valid URLs (All)}}
  & \multicolumn{4}{c}{\textbf{OADS URLs}} \\
\cmidrule(lr){2-7} \cmidrule(l){8-11}
  & \textbf{\#V}
  & \textbf{P [95\% CI]}
  & \textbf{R [95\% CI]}
  & \textbf{F1 [95\% CI]}
  & \textbf{$p$-value}
  & \textbf{Sig.}
  & \textbf{\#VO}
  & \textbf{R [95\% CI]}
  & \textbf{$p$-value}
  & \textbf{Sig.} \\
\midrule
\multicolumn{11}{l}{\textit{Single formats}} \\
  \quad \textbf{TEXTWAL (T)} & 1,887 & 0.60 [0.51,0.66] & 0.81 [0.75,0.85] & \textbf{0.69 [0.61,0.74]} & - & \textit{-} & 361 & 0.74 [0.68,0.80] & $<0.001$ & $^{***}$ \\
  \quad Markdown (M) & 1,474 & 0.69 [0.64,0.73] & 0.63 [0.53,0.73] & 0.66 [0.60,0.71] & 0.309 & \textit{ns} & 396 & \textbf{0.81 [0.71,0.89]} & - & \textit{-} \\
  \quad LaTeX (L) & 1,174 & 0.74 [0.67,0.80] & 0.50 [0.38,0.63] & 0.60 [0.49,0.69] & $<0.001$ & $^{***}$ & 304 & 0.62 [0.52,0.72] & $<0.001$ & $^{***}$ \\
  \quad HTML (H) & 827 & 0.49 [0.33,0.64] & 0.35 [0.23,0.49] & 0.41 [0.27,0.55] & $<0.001$ & $^{***}$ & 174 & 0.36 [0.26,0.46] & $<0.001$ & $^{***}$ \\
  \quad IMAGE (I) & 789 & 0.11 [0.08,0.15] & 0.34 [0.26,0.44] & 0.17 [0.13,0.21] & $<0.001$ & $^{***}$ & 313 & 0.64 [0.54,0.74] & 0.010 & $^{**}$ \\
  \quad XML (X) & 670 & 0.69 [0.63,0.74] & 0.29 [0.22,0.38] & 0.41 [0.33,0.50] & $<0.001$ & $^{***}$ & 206 & 0.42 [0.33,0.52] & $<0.001$ & $^{***}$ \\
\midrule
\multicolumn{11}{l}{\textit{Two formats}} \\
  \quad T + M & 2,098 & 0.55 [0.48,0.62] & 0.90 [0.86,0.92] & 0.68 [0.62,0.74] & $<0.001$ & $^{***}$ & 455 & \textbf{0.93 [0.90,0.96]} & 0.005 & $^{**}$ \\
  \quad \textbf{T + L} & 2,059 & 0.57 [0.50,0.63] & 0.88 [0.84,0.91] & \textbf{0.69 [0.63,0.75]} & $<0.001$ & $^{***}$ & 423 & 0.87 [0.81,0.91] & 1.000 & \textit{ns} \\
  \quad T + H & 2,049 & 0.49 [0.42,0.55] & 0.88 [0.84,0.91] & 0.63 [0.57,0.68] & $<0.001$ & $^{***}$ & 406 & 0.83 [0.77,0.88] & 0.025 & $^{*}$ \\
  \quad \textbf{T + X} & 2,035 & 0.57 [0.49,0.63] & 0.87 [0.83,0.90] & \textbf{0.69 [0.62,0.74]} & $<0.001$ & $^{***}$ & 424 & 0.87 [0.82,0.91] & 1.000 & \textit{ns} \\
\midrule
\multicolumn{11}{l}{\textit{Three formats}} \\
  \quad T + H + M & 2,174 & 0.46 [0.40,0.52] & 0.93 [0.90,0.95] & 0.62 [0.56,0.67] & $<0.001$ & $^{***}$ & 460 & 0.94 [0.91,0.97] & 0.003 & $^{**}$ \\
  
  \quad T + I + M & 2,130 & 0.21 [0.16,0.27] & 0.91 [0.88,0.93] & 0.34 [0.27,0.41] & $<0.001$ & $^{***}$ & 467 & \textbf{0.96 [0.93,0.98]} & $<0.001$ & $^{***}$ \\
  
  \quad \textbf{T + L + X}  & 2,107 & 0.54 [0.47,0.60] & 0.90 [0.86,0.93] & \textbf{0.68 [0.61,0.73]} & $<0.001$ & $^{***}$ & 448 & 0.92 [0.88,0.95] & 0.837 & \textit{ns} \\
  \quad L + H + X \dag & 1,576 & 0.51 [0.43,0.58] & 0.67 [0.55,0.78] & 0.58 [0.49,0.66] & 0.572 & \textit{ns} & 384 & 0.79 [0.70,0.87] & 1.000 & \textit{ns} \\
\midrule
\multicolumn{11}{l}{\textit{Four formats}} \\
  \quad T + L + H + M & 2,202 & 0.44 [0.39,0.50] & 0.94 [0.92,0.96] & 0.60 [0.55,0.65] & $<0.001$ & $^{***}$ & 469 & \textbf{0.96 [0.93,0.98]} & $<0.001$ & $^{***}$ \\
  \quad T + L + H + X & 2,169 & 0.45 [0.39,0.51] & 0.93 [0.90,0.95] & 0.61 [0.55,0.66] & $<0.001$ & $^{***}$ & 453 & 0.93 [0.89,0.96] & 0.692 & \textit{ns} \\
  \quad \textbf{T + L + X + M} & 2,154 & 0.50 [0.43,0.56] & 0.92 [0.89,0.94] & \textbf{0.65 [0.58,0.70]} & $<0.001$ & $^{***}$ & 467 & 0.96 [0.93,0.98] & $<0.001$ & $^{***}$ \\
\midrule
\multicolumn{11}{l}{\textit{Five formats}} \\
  \quad T + L + H + I + M & 2,222 & 0.20 [0.15,0.24] & 0.95 [0.93,0.97] & 0.33 [0.26,0.39] & $<0.001$ & $^{***}$ & 477 & \textbf{0.98 [0.96,0.99]} & $<0.001$ & $^{***}$ \\
  \quad \textbf{T + L + H + X + M} & 2,213 & 0.42 [0.37,0.47] & 0.95 [0.93,0.96] & \textbf{0.58 [0.53,0.63]} & $<0.001$ & $^{***}$ & 471 & 0.96 [0.94,0.99] & $<0.001$ & $^{***}$ \\
\midrule
\multicolumn{11}{l}{\textit{All formats}} \\
  \quad T + L + H + X + M + I & 2,224 & 0.19 [0.15,0.24] & 0.95 [0.93,0.97] & \textbf{0.32 [0.26,0.38]} & $<0.001$ & $^{***}$ & 477 & \textbf{0.98 [0.96,0.99]} & $<0.001$ & $^{***}$ \\
\bottomrule
\end{tabular}
}
\end{table*}

\subsubsection{Statistical Significance of Format Differences}
\label{sec:results:statistics}

\paragraph{Friedman test: global format differences.}
The Friedman test is highly significant for both metrics: All-URL Recall ($\chi^2(5)=232.49$, $p<0.001$, $n=200$) and OADS Recall ($\chi^2(5)=129.29$, $p<0.001$, $n=122$). This confirms that format choice materially affects performance and supports pairwise testing. 

Next, we compare each of the 62 format combinations against the strongest single-format baseline using Wilcoxon signed-rank tests. The baselines are TEXTWAL for All-URL Recall ($R = 0.81$) and Markdown for OADS Recall ($R = 0.81$), the top-performing single formats for their respective scenarios.

\paragraph{Single-format pairwise comparisons.}
For All-URL Recall, only Markdown is not significantly different from TEXTWAL ($p=0.309$); All other formats underperform TEXTWAL with statistical significance ($p<0.001$). For OADS Recall, all non-Markdown formats differ significantly from the Markdown baseline (TEXTWAL, LaTeX, HTML, XML: $p<0.001$; IMAGE: $p<0.01$), indicating Markdown uniquely achieves top per-paper OADS coverage.

\paragraph{Multi-format combination comparisons.}
For All-URL Recall, all 31 combinations containing TEXTWAL significantly outperform TEXTWAL alone ($p<0.001$), indicating consistent gains from complementary formats. In contrast, combinations without TEXTWAL rarely improve performance; those that do are uniformly worse than TEXTWAL, while most other combinations are statistically indistinguishable, suggesting similar single-source performance ceilings. For OADS Recall, 43 of 62 combinations significantly exceed Markdown, while the remaining 19 (mostly lacking Markdown or combining TEXTWAL with limited additional formats) do not, confirming Markdown's central role in OADS extraction.


\subsection{Analysis of URL extraction performance }
\label{sec:results:domain-coverage}
Coverage analysis across the top URL domains and protocols reveals systematic strengths and weaknesses of different extraction approaches.

\begin{figure}[ht]
  \centering
  \setlength{\fboxsep}{2pt}
  \setlength{\fboxrule}{0pt}
  \fbox{\includegraphics[trim={10pt 15pt 10pt 10pt}, clip, width=0.95\linewidth]{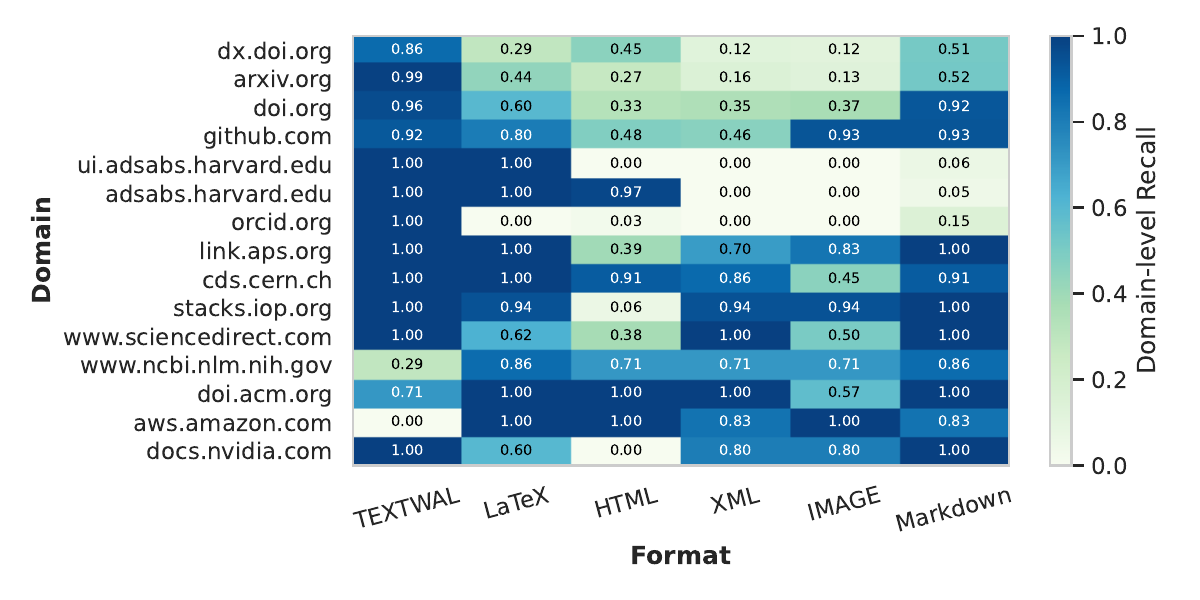}}
  \caption{\footnotesize Domain-level extraction recall across six file formats. The heatmap displays the recall for the top 15 domains identified in our corpus.}
  \label{fig:domain_coverage}
\end{figure}

\begin{figure}[ht]
  \centering
  \setlength{\fboxsep}{2pt}
  \setlength{\fboxrule}{0pt}
  \fbox{\includegraphics[trim={10pt 15pt 25pt 12pt}, clip, width=0.75\linewidth]{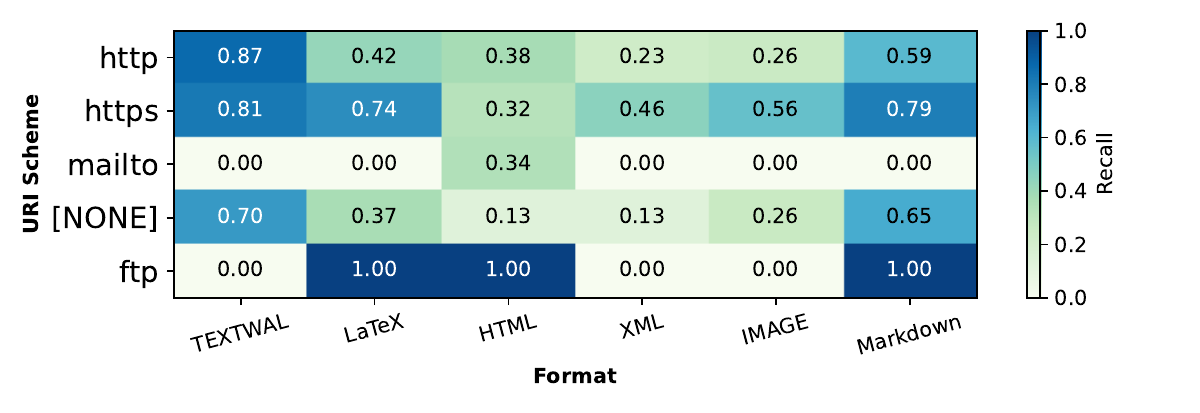}}
  \caption{\footnotesize Scheme-level extraction recall across six file formats. [NONE] denotes URIs without an explicit scheme.}
  \label{fig:protocol_coverage}
\end{figure}

The 15 most frequently occurring domains in our ground truth are predominantly from academic and scientific infrastructure and the coverage patterns diverge substantially by domain and format (Figure~\ref{fig:domain_coverage}). TEXTWAL achieves the highest recall for \texttt{arxiv.org} (0.99) and near-perfect coverage across DOI-based domains, as well as for physics publisher and library portals such as \texttt{link.aps.org}, \texttt{cds.cern.ch}, and \texttt{stacks.iop.org}. Notably, \texttt{orcid.org} is almost exclusively captured by TEXTWAL, with all other formats scoring near zero, indicating that ORCIDs appear predominantly as plain-text strings rather than hyperlinks in scholarly source files.

In contrast, URLs with the domain \texttt{aws.amazon.com} present the inverse pattern: TEXTWAL completely fails, while LaTeX, HTML, and IMAGE achieve perfect recall and XML and Markdown score 0.83. This suggests that AWS documentation links are primarily encoded as hyperlinks in source markup but are invisible or partially extracted (broken URLs) via TEXTWAL. For \texttt{github.com}, TEXTWAL, IMAGE, and Markdown yield the strongest recall, while HTML and XML underperform, likely due to complexities introduced during LaTeXML and GROBID conversions. Despite the strong performance of IMAGE for GitHub URLs, these results suggest that the effectiveness of visual extraction varies across websites rather than providing a uniform advantage.

The scheme-level analysis (Figure~\ref{fig:protocol_coverage}) reveals further format-specific behavior. \texttt{http} URLs are best captured by TEXTWAL and Markdown, with XML and IMAGE performing poorly. Schemeless URLs are best handled by TEXTWAL (0.70) and Markdown (0.65), with HTML and XML nearly missing them entirely, consistent with their dependence on explicit scheme markers for parsing. \texttt{ftp} URLs exhibit a stark split: LaTeX, HTML, and Markdown achieve perfect recall (1.00), while TEXTWAL, XML, and IMAGE completely miss them, suggesting that FTP links are reliably encoded in source markup but are not recovered by text-based or visual approaches.

\rr{This domain and scheme analysis provides guidance for selecting extraction formats based on resource type. TEXTWAL performs robustly across text-heavy scholarly collections, including arXiv- and DOI-centric sources, while LaTeX and HTML recover more URLs from structured or web-oriented representations. Combining TEXTWAL with Markdown or LaTeX further improves coverage across common web schemes such as \texttt{http}, \texttt{https}, and \texttt{ftp}, although \texttt{mailto} links and identifiers such as \texttt{orcid.org} may require additional handling. These patterns are consistent with our URL complexity taxonomy: text-based formats handle spatial complexity well (e.g., split-line URLs), whereas annotation-layer access is essential for representational complexity, \rr{explaining TEXTWAL's substantially higher recovery of annotation-only URLs (87.0\% vs.\ 44.6\% for Markdown and below 37\% for all other formats).}}


\section{Temporal Trends in arXiv URLs}
\label{sec:temporal}

\subsection{Analysis Pipeline}
\label{sec:temporal:setup}

\begin{figure}[h]
  \centering
  \footnotesize
  \setlength{\fboxsep}{2pt}\setlength{\fboxrule}{0pt}
  \fbox{\includegraphics[trim={20pt 10pt 20pt 12pt}, clip, width=0.85\linewidth]{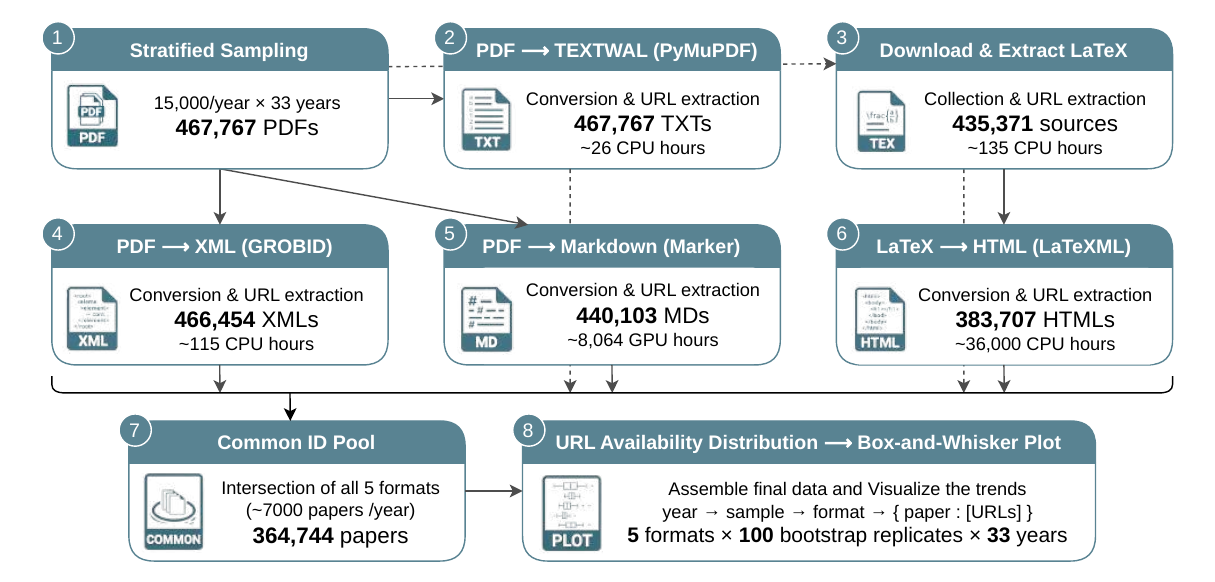}}
  
  \caption{\footnotesize Longitudinal extraction pipeline for the full arXiv corpus
           (1992--2024) across five document formats.
           Starting from 467,767 PDFs, successive format-specific conversion steps yield per-format pools; their intersection produces a common pool of ${\sim}$360k papers used for all comparisons. 100 bootstrap samples of 1,000 papers per year are drawn
           from the common pool, yielding $100 \times 33 = 3300$ year--sample combinations for bootstrap-based
           visualization.}
  \label{fig:longitudinal_pipeline}
\end{figure}

To assess whether format-specific coverage gaps observed in the 200-paper benchmark
generalize across the full arXiv corpus, we designed a longitudinal extraction pipeline
spanning 33 years (1992--2024), illustrated in Figure~\ref{fig:longitudinal_pipeline}.
Starting from a stratified random sample of up to 15,000 papers per year, we obtained an initial pool of 467,767 PDFs. These were processed in parallel through five format-specific pathways:

\begin{itemize}[leftmargin=6mm]
  \item \textbf{TEXTWAL} (PyMuPDF): applied to all 467,767 PDFs, producing one TXT
    file per paper in approximately 26 CPU-hours;
  \item \textbf{LaTeX}: source archives were retrieved and URL-extracted for 435,371
    papers (${\sim}$135 CPU-hours);
  \item \textbf{XML} (GROBID): 466,454 PDFs converted to TEI-XML and URL-extracted
    (${\sim}$115 CPU-hours);
  \item \textbf{HTML} (LaTeXML): available LaTeX sources converted to HTML for
    383,707 papers (${\sim}$36,000 CPU-hours; approximately 15 days elapsed
    across 100 parallel CPU processes);
  \item \textbf{Markdown} (Marker): 440,103 PDFs converted to Markdown
    (${\sim}$8,064 GPU-hours on NVIDIA A100s; approximately 30 days elapsed
    across 12 parallel GPU jobs).
\end{itemize}

Restricting to papers for which all five formats produced successful, non-empty output yields a common pool of 364,744 papers (${\sim}$7,000 per year) across all 33 years, ensuring every subsequent format comparison is made on identical paper sets. \rr{We publicly release the URLs extracted from all formats and make the resulting common-paper corpus ($\sim$360k) available via Hugging Face\footnote{\url{https://huggingface.co/datasets/rochanaro/hf-arxiv-url-bench/tree/main/arxiv-mini-corpus}}, enabling reproducibility and further large-scale studies.}

From this common pool, for each year we repeatedly sampled 1,000 papers
with replacement and applied format-specific URL extraction. We repeated this procedure for 100 bootstrap replicates per year \cite{Pattengale2010-du}, which we found sufficient for stable quartile and whisker summaries for visualization, while keeping the computation tractable across 33 yearly strata. The resulting distributions are summarized as box plots per format in Figure \ref{fig:url_temporal_trends} (\rro{Here we measure extraction \emph{yield}, not \emph{accuracy}, as ground truth at this scale is infeasible).}


\subsection{Overall URL Usage Trends}
\label{sec:temporal:trends}

\begin{figure*}[h]
  \centering
  \setlength{\fboxsep}{0.5pt}\setlength{\fboxrule}{0.0pt}
  \fbox{\includegraphics[trim=1 1 1 1, clip, width=1.0\linewidth]{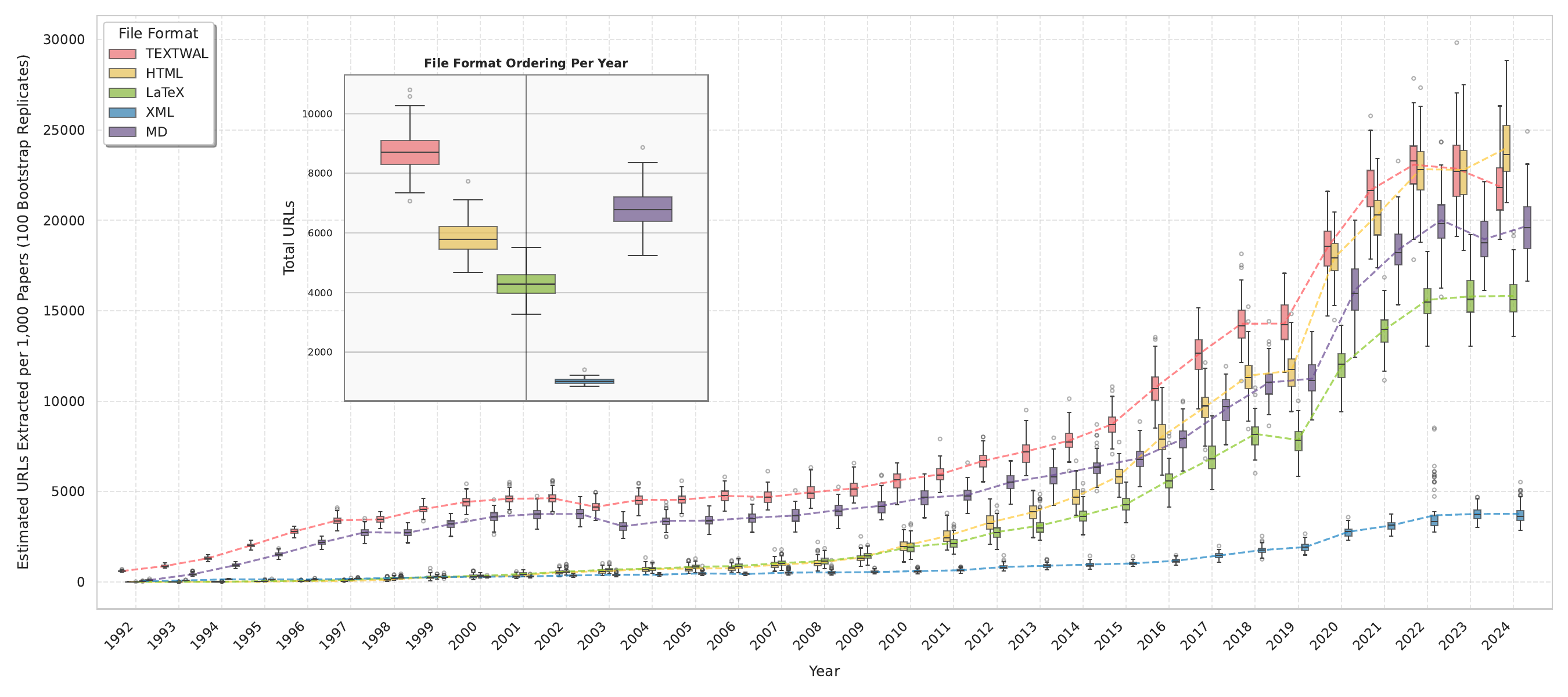}}
  \caption{Temporal distribution of extracted URLs from arXiv papers (1992--2024). Whisker plot shows mean and 95\% bootstrap CI per year from 100 random draws of 1,000 papers from the common five-format pool, reported separately for each of the five formats. Inset: Expanded view of a single year (2015), illustrating the horizontal offsets used to display the five file formats within each yearly bin.}
  \label{fig:url_temporal_trends}
\end{figure*}

URL usage in arXiv papers grows steadily from 1992 to around 2015, then increases sharply to a peak in 2022 before stabilizing in 2023--2024. Extraction yield differs substantially by format, revealing a consistent hierarchy throughout the 33-year period (Figure~\ref{fig:url_temporal_trends}). The TEXTWAL format yields the most URLs in every year, followed by Markdown and HTML, while LaTeX and especially XML recover substantially fewer links.

In 2015, TEXTWAL extracts approximately 8.8 URLs per paper, compared with 6.8 for Markdown, 5.8 for HTML, 4.5 for LaTeX, and 0.9 for XML. By 2022, these increase to approximately 28, 24, 26, 16, and 3 URLs per paper, respectively.\rr{The sharp improvement in HTML URL extraction after 2015 plausibly coincides with the maturation of LaTeXML-based conversion tools, which underwent substantial improvements in HTML conversion fidelity and package support during 2013–2015 \cite{ginev2014ebooksgraphicsLaTeXml, miller2015strategiesparallelmarkup}.} A noticeable dip is visible across all formats around 2019, suggesting a corpus-wide shift in citation or resource-linking practices rather than a format-specific extraction effect. A second decline occurs in 2023, followed by partial recovery in 2024. The persistent advantage of TEXTWAL, Markdown, and HTML reflects their ability to recover hyperlinks exposed through PDFs and compiled documents, whereas raw LaTeX can obscure links behind macros. The strong growth after 2015 coincides with increasing citation of code repositories (e.g., GitHub), model hubs, and research data archives, making modern papers substantially more URL-dense.


\section{Discussion}
\label{sec:discussion}

\subsection{Format and Approach Recommendations}
\label{sec:discussion:recommendations}
The decision framework in Table~\ref{tab:recommendations} summarizes how source availability (PDF and LaTeX), computational budget, and target objective jointly determine optimal extraction strategies \rro{(the framework is based on LaTeX-native arXiv papers with source files available; extraction behavior may differ for PDF-only or Word-derived submissions)}. We explicitly account for the computational cost of each pipeline. \rr{Our notion of computational budget is informed by the large-scale longitudinal analysis (Section \ref{sec:temporal}), which measured the CPU/GPU resources and processing time required to convert and extract URLs from more than 467k arXiv papers}. Overall, the table shows that higher recall and OADS coverage are consistently achieved through multi-format combinations, whereas TEXTWAL-centered pipelines provide a strong low-cost input format when only PDFs are available.

\vspace{-0.3cm}

\begin{table}[h]
\centering
\footnotesize

\caption{Recommended file formats for URL extraction across different practitioner scenarios (prioritizing recall). 
Comput. Cost denotes the relative computational cost of each pipeline, estimated from empirical CPU/GPU resource usage and processing time measured in Section~\ref{sec:temporal:setup} (Low: CPU-bound, Medium: Heavy CPU/Shared GPU, High: Dedicated GPU clusters).}

\label{tab:recommendations}
\begin{tabular}{p{0.17\columnwidth}p{0.10\columnwidth}p{0.48\columnwidth}p{0.14\columnwidth}}
\toprule
\textbf{Scenario} & \textbf{Availability} & \textbf{Format(s)} & \textbf{Comput. Cost} \\
\midrule
Unconstrained
  & PDF, LaTeX
  & TEXTWAL + LaTeX + HTML + XML + Markdown + IMAGE
  & High \\
\addlinespace
Unconstrained
  & PDF
  & TEXTWAL + XML + Markdown
  & High \\
\addlinespace
Resource-constrained
  & PDF, LaTeX
  & TEXTWAL + LaTeX + HTML + XML
  & Medium \\
\addlinespace
Resource-constrained
  & PDF, LaTeX
  & TEXTWAL + LaTeX
  & Low \\
\addlinespace
OADS-priority
  & PDF, LaTeX
  & TEXTWAL + LaTeX + HTML + XML + Markdown
  & High \\
\addlinespace
Single-format
  & PDF
  & TEXTWAL
  & Low \\
\bottomrule
\end{tabular}
\end{table}

\vspace{-4mm}

\subsection{Format-Specific Failure Modes}
\label{sec:discussion:error-analysis}
\rr{We analyze format-specific failure modes across spatial, representational, lexicographical, and document-region dimensions (Table~\ref{tab:error_summary} and Appendix \ref{sec:discussion:error:summary}). 
TEXTWAL exhibits comparatively low miss rates, with failures mainly arising from embedded references and footnotes, whereas HTML, XML, LaTeX, Markdown, and VLM are dominated by hidden annotation layers, reaching 97.8\% for XML. Lexicographical failures primarily involve URLs containing explicit schemes (e.g., \texttt{http}, \texttt{ftp}) and complex path/query components. Across all formats, references remain the most error-prone document region, with misses reaching 70.6\% for VLM, while TEXTWAL limits reference misses to 14.1\%.}

\subsection{Implications for Large-Scale OADS Preservation}
\label{sec:temporal:implications}

\rr{Since 2015, OADS platforms such as GitHub, HuggingFace, Zenodo, figshare, and OSF have become the primary driver of arXiv's growing URL density (Figure~\ref{fig:url_temporal_trends}), centralizing scholarly links in a small number of external infrastructures. Extrapolating from our benchmark ($\sim$21\% OADS URLs) suggests 4M--6M OADS URLs across arXiv, with roughly 0.8M--1.2M likely to become inaccessible within 3--5 years due to link rot \cite{Lemaire2025-kb}. Our computational cost analysis (Figure~\ref{fig:longitudinal_pipeline}) further shows that large-scale OADS URL extraction is computationally feasible, requiring roughly 1--2 months of processing for the full arXiv corpus.}

\begin{table}[h]
\centering
\caption{Dominant failure modes across extraction formats. Percentages denote the proportion of missed URLs associated with the primary failure mode(s) in each format.}
\label{tab:error_summary}
\footnotesize
\begin{tabular}{p{0.12\linewidth}p{0.18\linewidth}p{0.28\linewidth}p{0.26\linewidth}}

\toprule
\textbf{Format} & \textbf{Spatial / Representation} & \textbf{Lexicographical} & \textbf{Most Affected Sections} \\
\midrule

TEXTWAL (PyMuPDF)
& Embedded footnotes (42.2\%)
& Scheme URLs (18.9\%); path/query URLs (13.4\%)
& Methodology (22.8\%), Title (43.8\%) \\

HTML
& Hidden annotations (63.1\%)
& Scheme URLs (64.2\%); path/query URLs (65.0\%)
& References (60.5\%), Title (91.4\%), Methodology (70.8\%) \\

XML
& Hidden annotations (97.8\%); footnotes (91.3\%)
& Scheme URLs (70.9\%); path/query URLs (70.9\%)
& Title (100\%), Methodology (82.5\%) \\

LaTeX
& Hidden annotations (67.0\%)
& Scheme URLs (49.5\%); path/query URLs (48.6\%)
& References (45.9\%), Title (97.8\%), Methodology (29.8\%) \\

Markdown
& Hidden annotations (55.4\%)
& Scheme URLs (36.9\%); path/query URLs (35.5\%)
& References (33.7\%), Title (93.5\%), Methodology (14.0\%) \\

VLM (Qwen)
& Hidden annotations (96.5\%)
& Scheme URLs (66.0\%)
& References (70.6\%), Title (97.8\%)\\

\bottomrule
\end{tabular}
\end{table}

\vspace{-4mm}
\subsection{Limitations}
\label{sec:discussion:limitations}

Our study, focused on arXiv papers, may not fully generalize to other repositories with different document structures. \rr{Furthermore, our analysis considers only papers for which all formats are available, meaning the underlying PDFs are predominantly LaTeX-generated; PDFs converted from other sources, such as Word documents, may exhibit different structural characteristics that limit the generalizability of our findings.} Finally, LLM results depend on specific model versions which can improve over time.

\vspace{-3.5mm} 
\section{Conclusion and Future Work}
\label{sec:conclusion}

In this paper, we construct a benchmark consisting of 2,338 manually annotated URLs from 200 arXiv papers spanning 1992--2024 and seven academic domains. We systematically evaluate URL extraction performance across six document formats—text with annotation layer (TEXTWAL), LaTeX, HTML, XML, Markdown, and PNG (IMAGE)—each associated with its respective extraction method. Our extensive experiments indicate that extraction quality strongly depends on format selection.
Among structured approaches, TEXTWAL+LaTeX provides the best balance of precision, recall, and cost effectiveness ($F1{=}0.69$; $R{=}0.88$), while combining TEXTWAL, LaTeX, HTML, and Markdown recovers 96\% OADS URLs. For PDF-only collections, TEXTWAL remains the strongest single-format baseline ($F1{=}0.69$; $R{=}0.81$). LLMs on text-based formats consistently underperform rule-based methods in recall and are prohibitively expensive for large-scale deployment. Our complexity analysis further shows that no single format dominates across all URL types, highlighting the value of complementary extraction strategies. 

Beyond extraction performance, our large-scale analysis of 467,767 arXiv papers across all URLs confirms the findings of our benchmark study. The observed trends highlight the significant discrepancy of URLs extracted based on different input formats. This large-scale evidence further supports the need for robust multi-format extraction pipelines to achieve reliable URL recovery at corpus scale. The benchmark, annotation guidelines, format decision framework, and the mini version of the arXiv corpus across all six formats provide a foundation for future research. 


We will extend this work toward building large-scale datasets of OADS URLs, scaling beyond arXiv. We will investigate URL extraction from PDFs converted from non-LaTeX sources such as Word documents, where layout and encoding properties may differ substantially.


\rro{
\section*{Acknowledgments}  
This work was partially supported by the IMLS grant (LG-256694-OLS-24). We thank Gabriel Vega Osborne for his assistance with the URL annotation process.
}

\bibliographystyle{unsrt}  
\bibliography{references}

@String{Computing = "Computing" }

@String{Computer = "{IEEE} Computer" }

@String{Academic = "Academic Press" }

@String{Springer = "Springer-Verlag" }

@misc{acm-rep-24,
author = {Obadage, Rochana R. and Rajtmajer, Sarah M. and Wu, Jian},
title = {SHORT: Can citations tell us about a paper's reproducibility? A case study of machine learning papers},
year = {2024},
isbn = {9798400705304},
publisher = {Association for Computing Machinery},
address = {New York, NY, USA},
url = {https://doi.org/10.1145/3641525.3663628},
booktitle = {Proceedings of the 2nd ACM Conference on Reproducibility and Replicability},
pages = {96–100},
numpages = {5},
location = {Rennes, France},
series = {ACM REP '24}
}

@INPROCEEDINGS{11363696,
  author={Obadage, Rochana R. and Rajtmajer, Sarah and Wu, Jian},
  booktitle={2025 ACM/IEEE Joint Conference on Digital Libraries (JCDL)}, 
  title={C{C}30k: A Citation Contexts Dataset for Reproducibility-Oriented Sentiment Analysis}, 
  year={2025},
  volume={},
  number={},
  pages={20-29},
  doi={10.1109/JCDL67857.2025.00013}}

@misc{PyMuPDF,
	author = {Julian Smith},
	title = {PyMuPDF},
	howpublished = {\url{https://pypi.org/project/PyMuPDF/}},
	year = {2016},
	note = {[Accessed 27-06-2025]},
}

@misc{GROBID,
    title = {GROBID},
    howpublished = {\url{https://github.com/kermitt2/grobid}},
    publisher = {GitHub},
    year = {2008--2025},
    archivePrefix = {swh},
    eprint = {1:dir:dab86b296e3c3216e2241968f0d63b68e8209d3c}
}

@misc{githubGitHubBrucemillerLaTeXML,
	author = {Bruce R. Miller, Deyan Ginev},
	title = {LaTeXML: A LaTeX to XML/HTML/MathML Converter},
	howpublished = {\url{https://github.com/brucemiller/LaTeXML}},
	year = {2004},
	note = {[Accessed 27-06-2025]},
}

@ARTICLE{url_decay,
  title    = "The continued problem of {URL} decay: an updated analysis of
              health care management journal citations",
  author   = "Howell, Susan and Burtis, Amber",
  journal  = "J Med Libr Assoc",
  volume   =  110,
  number   =  4,
  pages    = "463--470",
  month    =  oct,
  year     =  2022,
  address  = "United States",
  language = "en"
}

@inproceedings{link_rot,
author = {Lakic, Viktor and Rossetto, Luca and Bernstein, Abraham},
title = {Link-Rot in Web-Sourced Multimedia Datasets},
year = {2023},
isbn = {978-3-031-27076-5},
publisher = {Springer-Verlag},
address = {Berlin, Heidelberg},
url = {https://doi.org/10.1007/978-3-031-27077-2_37},
doi = {10.1007/978-3-031-27077-2_37},
booktitle = {MultiMedia Modeling: 29th International Conference, MMM 2023, Bergen, Norway, January 9–12, 2023, Proceedings, Part I},
pages = {476–488},
numpages = {13},
location = {Bergen, Norway}
}

@misc{arxiv,
  author       = {{arXiv}},
  title        = {{arXiv.org: e-Print archive for Physics, Mathematics, Computer Science, Quantitative Biology, Quantitative Finance, and Statistics}},
  year         = {2026},
  howpublished = {\url{https://arxiv.org}},
  note         = {Accessed: 2026-04-22}
}

@inproceedings{david2025github,
author = {Calano, David and Nelson, Michael and Weigle, Michele},
title = {GitHub Repository Complexity Leads to Diminished Web Archive Availability},
year = {2025},
isbn = {9798400714832},
publisher = {Association for Computing Machinery},
address = {New York, NY, USA},
url = {https://doi.org/10.1145/3717867.3717920},
doi = {10.1145/3717867.3717920},
booktitle = {Proceedings of the 17th ACM Web Science Conference 2025},
pages = {449–459},
numpages = {11},
location = {
},
series = {Websci '25}
}

@article{klein2014justkeepingtrack,
    doi = {10.1371/journal.pone.0115253},
    author = {Klein, Martin AND Van de Sompel, Herbert AND Sanderson, Robert AND Shankar, Harihar AND Balakireva, Lyudmila AND Zhou, Ke AND Tobin, Richard},
    journal = {PLOS ONE},
    publisher = {Public Library of Science},
    title = {Scholarly Context Not Found: One in Five Articles Suffers from Reference Rot},
    year = {2014},
    month = {12},
    volume = {9},
    url = {https://doi.org/10.1371/journal.pone.0115253},
    pages = {1-39},
    number = {12},
}

@ARTICLE{Hennessey2013,
  title    = "A cross disciplinary study of link decay and the effectiveness of
              mitigation techniques",
  author   = "Hennessey, Jason and Ge, Steven Xijin",
  journal  = "BMC Bioinformatics",
  volume   =  14,
  number   =  14,
  pages    = "S5",
  month    =  oct,
  year     =  2013
}

@misc{kenny-icdar-2023,
author="Ajayi, Kehinde
and Choudhury, Muntabir Hasan
and Rajtmajer, Sarah M.
and Wu, Jian",
editor="Fink, Gernot A.
and Jain, Rajiv
and Kise, Koichi
and Zanibbi, Richard",
title="A Study on Reproducibility and Replicability of Table Structure Recognition Methods",
booktitle="Document Analysis and Recognition - ICDAR 2023",
year="2023",
publisher="Springer Nature Switzerland",
address="Cham",
pages="3--19",
isbn="978-3-031-41679-8",
howpublished = {\url{https://doi.org/10.1007/978-3-031-41679-8_1}}
}

@inproceedings{salsabil2025context,
  title={Context-Based URL Classification for Open Access Datasets and Software in Scholarly Documents},
  author={Salsabil, Lamia and Obadage, Rochana R and Banerjee, Bipasha and Abeysinghe, Yasasi and Alam, Sawood and F{\"a}rber, Michael and Ingram, William and Fox, Edward and Wu, Jian},
  booktitle={2025 ACM/IEEE Joint Conference on Digital Libraries (JCDL)},
  pages={197--206},
  year={2025},
  organization={IEEE}
}

@misc{ginsparg2011yearsagotoday,
      title={It was twenty years ago today ...}, 
      author={Paul Ginsparg},
      year={2011},
      eprint={1108.2700},
      archivePrefix={arXiv},
      primaryClass={cs.DL},
      url={https://arxiv.org/abs/1108.2700}, 
}

@BOOK{national2019reproducibility,
  author    = "National Academies of Sciences, Engineering and Medicine",
  title     = "Reproducibility and Replicability in Science",
  isbn      = "978-0-309-48616-3",
  doi       = "10.17226/25303",
  url       = "https://nap.nationalacademies.org/catalog/25303/reproducibility-and-replicability-in-science",
  year      = 2019,
  publisher = "The National Academies Press",
  address   = "Washington, DC"
}

@InProceedings{escamilla2023itsjustgithubidentifying,
author="Escamilla, Emily
and Salsabil, Lamia
and Klein, Martin
and Wu, Jian
and Weigle, Michele C.
and Nelson, Michael L.",
editor="Alonso, Omar
and Cousijn, Helena
and Silvello, Gianmaria
and Marrero, M{\'o}nica
and Teixeira Lopes, Carla
and Marchesin, Stefano",
title="It's Not Just GitHub: Identifying Data and Software Sources Included in Publications",
booktitle="Linking Theory and Practice of Digital Libraries",
year="2023",
publisher="Springer Nature Switzerland",
address="Cham",
pages="195--206",
isbn="978-3-031-43849-3"
}

@inproceedings{s2orc,
  title={{S2ORC}: The Semantic Scholar Open Research Corpus},
  author={Lo, Kyle and Wang, Lucy Lu and Neumann, Mark and Kinney, Rodney and Weld, Daniel S},
  booktitle={Proceedings of the 58th Annual Meeting of the Association for Computational Linguistics (ACL)},
  pages={4969--4983},
  year={2020}
}

@inproceedings{lopez2009grobid,
  title={{GROBID}: Combining Automatic Bibliographic Data Recognition and Term Extraction for Scholarship Publications},
  author={Lopez, Pierre},
  booktitle={Proceedings of the 13th European Conference on Digital Libraries (ECDL)},
  pages={473--474},
  year={2009}
}

@misc{Datalabtomarker,
	author = {Datalab},
	title = {Marker: Convert {P}{D}{F} to markdown + {J}{S}{O}{N} quickly with high accuracy --- github.com},
	howpublished = {\url{https://github.com/datalab-to/marker}},
	year = {2024},
	note = {[Accessed 16-06-2026]},
}

@article{placeholder2024qwen,
  title={Qwen2-VL: Enhancing Vision-Language Model's Perception of the World at Any Resolution},
  author={Wang, Peng and Bai, Shuai and Tan, Sinan and Wang, Shijie and Fan, Zhihao and Bai, Jinze and Chen, Keqin and Liu, Xuejing and Wang, Jialin and Ge, Wenbin and Fan, Yang and Dang, Kai and Du, Mengfei and Ren, Xuancheng and Men, Rui and Liu, Dayiheng and Zhou, Chang and Zhou, Jingren and Lin, Junyang},
  journal={arXiv preprint arXiv:2409.12191},
  year={2024}
}

@misc{placeholder2024minicpm,
      title={MiniCPM: Unveiling the Potential of Small Language Models with Scalable Training Strategies}, 
      author={Shengding Hu and Yuge Tu and Xu Han and Chaoqun He and Ganqu Cui and Xiang Long and Zhi Zheng and Yewei Fang and Yuxiang Huang and Weilin Zhao and Xinrong Zhang and Zheng Leng Thai and Kaihuo Zhang and Chongyi Wang and Yuan Yao and Chenyang Zhao and Jie Zhou and Jie Cai and Zhongwu Zhai and Ning Ding and Chao Jia and Guoyang Zeng and Dahai Li and Zhiyuan Liu and Maosong Sun},
      year={2024},
      eprint={2404.06395},
      archivePrefix={arXiv},
      primaryClass={cs.CL},
      url={https://arxiv.org/abs/2404.06395}, 
}

@misc{placeholder2024deepseek,
      title={DeepSeek-VL: Towards Real-World Vision-Language Understanding}, 
      author={Haoyu Lu and Wen Liu and Bo Zhang and Bingxuan Wang and Kai Dong and Bo Liu and Jingxiang Sun and Tongzheng Ren and Zhuoshu Li and Hao Yang and Yaofeng Sun and Chengqi Deng and Hanwei Xu and Zhenda Xie and Chong Ruan},
      year={2024},
      eprint={2403.05525},
      archivePrefix={arXiv},
      primaryClass={cs.AI},
      url={https://arxiv.org/abs/2403.05525}, 
}

@misc{PyPDF,
	author = {Mathieu Fenniak, Martin Thoma},
	title = {PyPDF --- pypi.org},
	howpublished = {\url{https://pypi.org/project/pypdf/}},
	year = {2025},
	note = {[Accessed 16-06-2026]},
}

@misc{pdfminer.six,
	author = {Yusuke Shinyama and Philippe Guglielmetti and Pieter Marsman},
	title = {{PDFMiner}: A tool for extracting information from PDF documents},
	howpublished = {\url{https://pypi.org/project/pdfminer.six/}},
	year = {2026},
	note = {[Accessed 16-06-2026]},
}

@misc{ncbi_pmc,
  author       = {{National Center for Biotechnology Information (NCBI)}},
  title        = {PubMed Central (PMC)},
  year         = {1988},
  address      = {Bethesda (MD): National Library of Medicine (US)},
  url          = {https://www.ncbi.nlm.nih.gov/pmc/},
  note         = {Accessed: 2026-04-23}
}

@misc{obadage2025robusturlextractionopen,
      title={Toward Robust URL Extraction for Open Science: A Study of arXiv File Formats and Temporal Trends}, 
      author={Rochana R. Obadage and Lamia Salsabil and Sawood Alam and Bipasha Banarjee and William A. Ingram and Edward A. Fox and Jian Wu},
      year={2025},
      eprint={2509.04759},
      archivePrefix={arXiv},
      primaryClass={cs.DL},
      url={https://arxiv.org/abs/2509.04759}, 
}

@misc{mole-2025,
      title={MOLE: Metadata Extraction and Validation in Scientific Papers Using LLMs}, 
      author={Zaid Alyafeai and Maged S. Al-Shaibani and Bernard Ghanem},
      year={2025},
      eprint={2505.19800},
      archivePrefix={arXiv},
      primaryClass={cs.CL},
      url={https://arxiv.org/abs/2505.19800}, 
}

@inproceedings{STATUS-Bench,
author = {Ukai, Mahiro and Kurita, Shuhei and Inoue, Nakamasa},
title = {STATUS Bench: A Rigorous Benchmark for Evaluating Object State Understanding in Vision-Language Models},
year = {2025},
isbn = {9798400720352},
publisher = {Association for Computing Machinery},
address = {New York, NY, USA},
url = {https://doi.org/10.1145/3746027.3755565},
doi = {10.1145/3746027.3755565},
booktitle = {Proceedings of the 33rd ACM International Conference on Multimedia},
pages = {4718–4727},
numpages = {10},
location = {Dublin, Ireland},
series = {MM '25}
}

@ARTICLE{Pattengale2010-du,
  title    = "How many bootstrap replicates are necessary?",
  author   = "Pattengale, Nicholas D and Alipour, Masoud and Bininda-Emonds,
              Olaf R P and Moret, Bernard M E and Stamatakis, Alexandros",
  journal  = "J Comput Biol",
  volume   =  17,
  number   =  3,
  pages    = "337--354",
  month    =  mar,
  year     =  2010,
  address  = "United States",
  language = "en"
}

@misc{ginev2014ebooksgraphicslatexml,
      title={E-books and Graphics with LaTeXML}, 
      author={Deyan Ginev and Bruce R. Miller and Silviu Oprea},
      year={2014},
      eprint={1404.6547},
      archivePrefix={arXiv},
      primaryClass={cs.DL},
      url={https://arxiv.org/abs/1404.6547}, 
}

@misc{miller2015strategiesparallelmarkup,
      title={Strategies for Parallel Markup}, 
      author={Bruce R. Miller},
      year={2015},
      eprint={1507.00524},
      archivePrefix={arXiv},
      primaryClass={cs.DL},
      url={https://arxiv.org/abs/1507.00524}, 
}

@misc{datastet,
	author = {Patrice Lopez, Luca Foppiano},
	title = {{G}it{H}ub - {D}ata{S}eer/datastet: {F}inding mentions and citations to named and implicit research datasets from within the academic literature --- github.com},
	howpublished = {\url{https://github.com/DataSeer/datastet}},
	year = {2023},
	note = {[Accessed 06-07-2026]},
}

@ARTICLE{Lemaire2025-kb,
  title    = "Web references are not eternal: time-trend and qualitative impact
              of the loss of access to online resources cited in peer-reviewed
              medical journals",
  author   = "Lemaire, Benjamin and Bauer, Fan{\'e}lie and Chaves Rodriguez,
              Elena and Ghesquiere, Jonathan and Radziejwoski, Amandine and
              Roth, Aur{\'e}lie and Boyer, Maud",
  journal  = "Curr Med Res Opin",
  volume   =  41,
  number   =  3,
  pages    = "543--548",
  month    =  mar,
  year     =  2025,
  address  = "England",
  language = "en"
}


\newpage
\clearpage

\appendix


\begin{center}
    {\fontsize{18.5pt}{10pt}\selectfont\bfseries Appendix}
\end{center}
\vspace{1em}

\section{RegEx Pattern for URL Extraction}
\label{app:regex}

The URL extraction pipeline relies on a comprehensive regular expression designed to capture a wide range of URL formats encountered in scholarly documents, including standard web URLs, protocol-specific resources, domain-only references, authenticated URLs, IP-based addresses, Unicode domain names, ports, and resource paths. 

\begin{tcolorbox}[
    colback=gray!3,
    colframe=black!75,
    title=\textbf{URL Extraction RegEx Pattern},
    sharp corners,
    breakable,
    boxrule=0.8pt,
    fonttitle=\small\sffamily
]
\begin{lstlisting}[
    language=Python,
    basicstyle=\ttfamily\scriptsize,
    breaklines=true,
    columns=fullflexible,
    showspaces=false,         % Removes visible space symbols in indentation
    showstringspaces=false,   % Removes visible space symbols inside the triple-quoted string
    keepspaces=true,          % Keeps your actual layout alignment intact
    label={lst:url_regex}
]
url_pattern = re.compile(r'''(?xi)
\b(?:
    (?:https?|ftp|file|data|javascript|mailto|tel|git|ssh|magnet)://
    | www\d{0,3}[.]
    | [a-z0-9.\-]+[.][a-z]{2,4}/
)
(?:\S+(?::\S*)?@)?
(?:
    (?!(?:10|127)(?:\.\d{1,3}){3})
    (?!(?:169\.254|192\.168)(?:\.\d{1,3}){2})
    (?!172\.(?:1[6-9]|2\d|3[0-1])(?:\.\d{1,3}){2})
    (?:[1-9]\d?|1\d\d|2[01]\d|22[0-3])
    (?:\.(?:1?\d{1,2}|2[0-4]\d|25[0-5])){2}
    (?:\.(?:[1-9]\d?|1\d\d|2[0-4]\d|25[0-4]))
|
    (?:
        (?:
            [a-z0-9\u00a1-\uffff]
            [a-z0-9\u00a1-\uffff_-]{0,62}
        )?
        [a-z0-9\u00a1-\uffff]\.
    )*
    (?:[a-z\u00a1-\uffff]{2,}\.?)
)
(?::\d{2,5})?
(?:[/?#][^\s]*)?
\b
''')
\end{lstlisting}
\end{tcolorbox}

The pattern supports multiple URL schemes (e.g., HTTP(S), FTP, SSH, Git, Mailto, and Magnet links), domain-only references, IPv4 addresses, Unicode hostnames, optional authentication credentials, port numbers, and arbitrary resource paths. Private and local network ranges are excluded to reduce false positives arising from non-public addresses embedded in documents.

\section{Three-Dimensional URL Complexity Taxonomy}
\label{app:url_complexities_taxonomy}

To support the fine-grained diagnosis of extraction failures discussed in the main text, this appendix details our three-dimensional taxonomy for characterizing URL complexities. The extraction and annotation of URLs in scholarly PDFs must account for a wide range of structural, physical, and representational challenges. Consequently, we categorize these challenges into three distinct dimensions—Lexicographical, Spatial, and Representational—providing a standardized framework for evaluating extraction performance.

\subsection{Lexicographical Complexity}
This dimension refers to the structural complexity of the URL based on its inherent components. 

\begin{itemize}[leftmargin=*]
    \item \textbf{URLs with schemes:} URLs containing standard protocol identifiers.
    \begin{tcolorbox}[colback=blue!5, colframe=blue!50!black, boxrule=0.5pt, left=5pt, top=2pt, bottom=2pt]
    \texttt{http://www.pardus.at} \\
    \texttt{https://github.com/ghif/drcn} 
    \end{tcolorbox}
    
    \item \textbf{URLs with subdomains:} URLs structured with additional domain prefixes.
    \begin{tcolorbox}[colback=blue!5, colframe=blue!50!black, boxrule=0.5pt, left=5pt, top=2pt, bottom=2pt]
    \texttt{http://infoscience.epfl.ch/record/204670} 
    \end{tcolorbox}

    \item \textbf{URLs with port numbers:} URLs explicitly defining a network port.
    \begin{tcolorbox}[colback=blue!5, colframe=blue!50!black, boxrule=0.5pt, left=5pt, top=2pt, bottom=2pt]
    \texttt{http://intranet.corp.com:1080} 
    \end{tcolorbox}

    \item \textbf{URLs with Unicode characters:} URLs containing non-ASCII characters.
    \begin{tcolorbox}[colback=blue!5, colframe=blue!50!black, boxrule=0.5pt, left=5pt, top=2pt, bottom=2pt]
    \texttt{es.wikipedia.org/wiki/Lincoln\_Díaz-Balart} 
    \end{tcolorbox}

    \item \textbf{URLs with paths, query strings, and fragments:} Complex URLs pointing to specific resources or tracking parameters.
    \begin{tcolorbox}[colback=blue!5, colframe=blue!50!black, boxrule=0.5pt, left=5pt, top=2pt, bottom=2pt]
    \texttt{http://cdsweb.cern.ch/search?p=LHCb-PUB...} 
    \end{tcolorbox}

    \item \textbf{URLs with escape characters:} URLs utilizing percent-encoding for spaces or special characters.
    \begin{tcolorbox}[colback=blue!5, colframe=blue!50!black, boxrule=0.5pt, left=5pt, top=2pt, bottom=2pt]
    \texttt{http://www.cse.ust.hk/\%7Eyinz/htl4ic.zip} 
    \end{tcolorbox}

    \item \textbf{URLs without scheme:} Bare domains without HTTP/HTTPS/FTP prefixes.
    \begin{tcolorbox}[colback=blue!5, colframe=blue!50!black, boxrule=0.5pt, left=5pt, top=2pt, bottom=2pt]
    \texttt{thelawdictionary.org} \quad | \quad \texttt{aaai.org} 
    \end{tcolorbox}

    \item \textbf{URLs with trailing slashes:} URLs terminating with a forward slash.
    \begin{tcolorbox}[colback=blue!5, colframe=blue!50!black, boxrule=0.5pt, left=5pt, top=2pt, bottom=2pt]
    \texttt{http://webhome.cs.uvic.ca/\textasciitilde dmaslov/} 
    \end{tcolorbox}
\end{itemize}

\subsection{Spatial Complexity}
This dimension refers to how the URL is physically represented and positioned within the layout of the document.

\begin{itemize}[leftmargin=*]
    \item \textbf{Split across lines or pages:} URLs broken across multiple lines or spanning two pages due to document formatting.
    \begin{tcolorbox}[colback=green!5, colframe=green!50!black, boxrule=0.5pt, left=5pt, top=2pt, bottom=2pt]
    \texttt{https://sencanada.ca/content/sen/Committee/402/} \\
    \texttt{popu/rep/rephealthjun09-e.pdf} 
    \end{tcolorbox}

    \item \textbf{Embedded in layout elements:} URLs situated in specific document regions such as footnotes, figure/table captions, or references.
    \begin{tcolorbox}[colback=green!5, colframe=green!50!black, boxrule=0.5pt, left=5pt, top=2pt, bottom=2pt]
    \textit{Caption Example:} "...An animated evolution of this network can be seen at \texttt{http://www.youtube.com/user/complexsystemsvienna}" 
    \end{tcolorbox}
\end{itemize}

\subsection{Representational Complexity}
This dimension addresses how the URL is technically stored, displayed, or obfuscated within the document's structure.

\begin{itemize}[leftmargin=*]
    \item \textbf{Hidden in annotation layers or metadata:} URLs stored in the PDF's clickable annotation layer or document metadata, but not necessarily visible as plain text.
    \begin{tcolorbox}[colback=red!5, colframe=red!50!black, boxrule=0.5pt, left=5pt, top=2pt, bottom=2pt]
    \textit{Clickable text "arXiv: 1605.01082" linking to:} \\
    \texttt{http://arxiv.org/abs/1605.01082} 
    \end{tcolorbox}

    \item \textbf{Embedded in visual media:} URLs that exist entirely as part of an image or encoded within a QR code.
    \begin{tcolorbox}[colback=red!5, colframe=red!50!black, boxrule=0.5pt, left=5pt, top=2pt, bottom=2pt]
    \textit{https://www.nutrient.io/blog/barcode-ocr/}
    \end{tcolorbox}

    \item \textbf{Shortened URLs that redirect:} URLs utilizing external shortening services that mask the true destination.
    \begin{tcolorbox}[colback=red!5, colframe=red!50!black, boxrule=0.5pt, left=5pt, top=2pt, bottom=2pt]
    \texttt{http://ow.ly/BqCf30jqffN} \quad | \quad \texttt{https://goo.gl/sX2Apb} 
    \end{tcolorbox}
\end{itemize}

\section{VLM and LLM Prompt Specifications}
\label{app:prompts}

This section details the exact prompt configurations utilized for the Vision-Language Model (VLM) and Large Language Model (LLM) extraction workflows.

\subsection{Vision-Language Model (VLM) Prompt Design}
\label{sec:app:vlm-prompt}
For page-image extraction across open-source VLMs, models were evaluated using a zero-shot prompt template.

\begin{tcolorbox}[
    colback=gray!3,
    colframe=blue!45!black,
    title=\textbf{VLM Page-Image Extraction Prompt},
    sharp corners,
    breakable,
    boxrule=0.8pt,
    fonttitle=\small\sffamily\bfseries
]
\begin{lstlisting}[
    basicstyle=\ttfamily\scriptsize,
    breaklines=true,
    columns=fullflexible,
    showspaces=false,
    showstringspaces=false,
    keepspaces=true
]
You are an expert URL extractor. Your task is to identify all URLs in the provided page image.

Rules:
1. Only output URLs. Do NOT include any extra text or commentary.
2. Each URL must be on a separate line.
3. Include URLs in headers, footers, tables, or broken across multiple lines.
4. Reconstruct any URLs broken across lines.
5. Ignore all other text on the page.
6. Do not summarize or explain anything.

Output example:
https://example.com/page
http://domain.org/path?query
www.site.net/index.html
\end{lstlisting}
\end{tcolorbox}

\subsection{Large Language Model (LLM) Prompt Design}
\label{sec:app:llm-prompt}
Text-based URL extraction from the compiled TEXTWAL content was processed via structured API calls configured at temperature zero. 

\begin{tcolorbox}[
    colback=gray!3,
    colframe=purple!45!black,
    title=\textbf{LLM TEXTWAL Extraction Prompt},
    sharp corners,
    breakable,
    boxrule=0.8pt,
    fonttitle=\small\sffamily\bfseries
]
\begin{lstlisting}[
    basicstyle=\ttfamily\scriptsize,
    breaklines=true,
    columns=fullflexible,
    showspaces=false,
    showstringspaces=false,
    keepspaces=true
]
You are an expert data extraction assistant. Your goal is to analyze the provided academic paper content and extract every unique URL and DOI link found within the text.

### Extraction Rules
1. **Scope**: Extract URLs from the main body, footnotes, references, metadata, and tables.
2. **Reconstruction**: If a URL is split across two lines or contains a hyphen due to text wrapping (e.g., `https://example.com/sub-` on one line and `directory` on the next), reconstruct it into a single, continuous string.
3. **DOIs**: Include Digital Object Identifiers. If a DOI is provided as a raw string (e.g., `10.1038/s41586-020-2012-7`), format it as a full URL: `https://doi.org/[DOI]`.
4. **Validation**: Ensure the links are complete. Remove trailing punctuation (like periods or closing parentheses) that are not part of the actual web address.
5. **De-duplication**: Provide only unique URLs.

### Output Format
Return ONLY a valid JSON array of strings. Do not include introductory text, explanations, or Markdown code blocks. 

Format: ["https://url1.com", "https://url2.org"]

### Content to Analyze
{content}
\end{lstlisting}
\end{tcolorbox}

\section{OADS URL Categorization Taxonomy}
\label{app:oads_taxonomy}

To support the fine-grained distribution metrics presented in the main text (Section~\ref{sec:dataset:oads}), this appendix details the definitions and contextual examples for the six-class Open Access Data and Software (OADS) taxonomy adapted from Salsabil et al. This taxonomy classifies ground-truth URLs based on their provenance, ownership, and functional role within scholarly work.

\begin{itemize}[leftmargin=*]
    \item \textbf{1. Third-Party Dataset (TPD):} This category includes URLs linking to datasets that are generated, managed, and distributed by entities independent of the data users or authors of the publications. In these scenarios, \textit{authors have used already available data}.
    \begin{tcolorbox}[colback=teal!5, colframe=teal!50!black, boxrule=0.5pt, left=5pt, top=2pt, bottom=2pt]
    \textit{Contextual Example:} ``The NPI dataset was downloaded from \texttt{https://openpsychometrics.org/}.''
    \end{tcolorbox}
    
    \item \textbf{2. Third-Party Software (TPS):} URLs in this category link to software, programming scripts, source code, or computational tools developed, maintained, and distributed by entities separate from the software users or authors of the publications.
    \begin{tcolorbox}[colback=teal!5, colframe=teal!50!black, boxrule=0.5pt, left=5pt, top=2pt, bottom=2pt]
    \textit{Contextual Example:} ``Venn diagrams were created using VENNY 2.0.2 (\texttt{\small http://bioinfogp.cnb.icsc.es/tools/venny/index.html}).''
    \end{tcolorbox}

    \item \textbf{3. Author-Provided Dataset (APD):} This category contains links to datasets featured, generated, or cited in the academic publication that are directly provided and hosted by the authors of those specific publications. In these cases, \textit{authors have prepared their own data}.
    \begin{tcolorbox}[colback=teal!5, colframe=teal!50!black, boxrule=0.5pt, left=5pt, top=2pt, bottom=2pt]
    \textit{Contextual Example:} ``Our dataset of C projects is publicly available at \texttt{https://utexas.box.com/icsme2014-practice}.''
    \end{tcolorbox}

    \item \textbf{4. Author-Provided Software (APS):} URLs in this category point directly to source code, specialized implementations, libraries, or operational tools developed and shared explicitly by the authors of the target publication.
    \begin{tcolorbox}[colback=teal!5, colframe=teal!50!black, boxrule=0.5pt, left=5pt, top=2pt, bottom=2pt]
    \textit{Contextual Example:} ``Our implementation is free software and can be found online (\texttt{https://github.com/SnippyHolloW/OpeningTech/}).''
    \end{tcolorbox}

    \item \textbf{5. Project:} This unified category encompasses complex repositories or landing hubs containing an integrated mix of both datasets and software, or supporting experimental frameworks.
    \begin{tcolorbox}[colback=teal!5, colframe=teal!50!black, boxrule=0.5pt, left=5pt, top=2pt, bottom=2pt]
    \textit{Contextual Example:} ``The source codes, datasets, and documents are released in our github repository (\texttt{http://github.com/graphbig/graphBIG}).''
    \end{tcolorbox}

    \item \textbf{6. General URL (GenURL):} A non-OADS baseline category capturing all administrative, informational, contextual, or baseline URLs that link to resources other than specialized research datasets, software frameworks, or digital project repositories.
    \begin{tcolorbox}[colback=teal!5, colframe=teal!50!black, boxrule=0.5pt, left=5pt, top=2pt, bottom=2pt]
    \textit{Contextual Example:} ``The study was approved by the french national bodies responsible for ethics and privacy, the ``commission nationale de l'informatique et des libert'' (cnil, \texttt{http://www.cnil.fr}) and the ``comit de protection des personnes'' (\texttt{http://www.cppsudest2.com/}) of the hospital.''
    \end{tcolorbox}
\end{itemize}

\begin{table}[h]
\centering
\caption{Prevalance of OADS URL categories in the annotated URLs.}
\label{tab:oads_distribution}
\begin{tabular}{lrr}
\toprule
\textbf{OADS URL Category} & \textbf{\# URLs} & \textbf{\%} \\
\midrule
Third Party Dataset (TPD)      & 146   & 6.24\%  \\
Third Party Software (TPS)     & 207   & 8.85\%  \\
Author Provided Dataset (APD)  & 19    & 0.81\%  \\
Author Provided Software (APS) & 16    & 0.68\%  \\
Project                        & 100   & 4.28\%  \\
General URL (GenURL, non-OADS) & 1,850 & 79.13\% \\
\midrule
\textbf{Total}        & 2,338 & 100.00\% \\
\bottomrule
\end{tabular}
\end{table}

\section{Statistical Testing}
\label{app:stats}

Since labeled ground truth was available for only 200 papers, all reported
performance metrics are accompanied by statistical uncertainty estimates.
We apply a three-tier inference framework to support principled comparison
across format variants and combinations.

\subsection{Statistical Unit and Paired Design}

The statistical unit for all tests is the individual paper (\texttt{arxiv\_id}).
For each paper we compute per-document precision, recall, and F1 scores for
every format variant and combination.
Sampling is conducted at the document level to preserve pairing across file
formats, ensuring that each format is evaluated on an identical set of documents.

\subsection{Bootstrap Confidence Intervals}

We estimate 95\% confidence intervals for precision, recall, and F1 via paired
bootstrap resampling.
For each method, we resample the 200-paper document set with replacement for
$B = 10{,}000$ iterations, recomputing the corpus-level metric in each iteration.
The 2.5th and 97.5th percentiles of the bootstrap distribution form the
reported 95\% CI.
Bootstrap CIs are reported for all entries in the main combination table
(Table~\ref{tab:combinations}) in the compact form
$\text{metric}\ [\text{lower},\, \text{upper}]$.
Paired bootstrap resampling preserves within-paper correlations across formats,
yielding valid CIs for union-combination methods.

\subsection{Friedman Test for Global Format Comparison}

To test whether any statistically significant difference exists in Recall
performance across the six single-format variants simultaneously, we apply a
Friedman test (non-parametric repeated-measures ANOVA).
The Friedman test treats per-paper Recall scores as the response variable and
format variant as the within-subject factor, and the test is applied
independently to two metrics: All-URL Recall and OADS Recall.
A significant result ($p < 0.05$) confirms that at least one format variant
systematically differs from the others, providing statistical justification for
the pairwise comparisons that follow.

We prefer the Friedman test over a parametric Repeated Measures ANOVA
(RM-ANOVA) for two reasons specific to our experimental structure.
First, RM-ANOVA assumes that the outcome variable is approximately normally
distributed within each condition, whereas per-paper recall values are strictly
bounded in $[0,\,1]$ and typically exhibit bimodal skew, with mass concentrated
near 0 and 1 rather than a symmetric distribution.
Second, because scores are compressed against the $[0,\,1]$ ceiling and floor,
arithmetic means and standard deviations become systematically distorted by
boundary effects; by ranking scores within each paper rather than using raw
values, the Friedman test is robust to both distributional violations and
boundary constraints.

\subsection{Wilcoxon Signed-Rank Tests for Pairwise Comparisons}

Following a significant Friedman result, we perform pairwise Wilcoxon
signed-rank tests on per-paper Recall scores, comparing each of the 62
remaining format combinations to the best-performing single-format baseline
for that metric.
This ensures that significance reflects improvement beyond the strongest
single-format result, rather than a weaker or arbitrary reference.
The Wilcoxon signed-rank test is preferred over a paired $t$-test because
per-paper recall distributions are non-normal, violating the normality
assumption on which the $t$-test's validity depends.
The signed-rank test is further preferred over an unpaired alternative such as
the Mann--Whitney U test because per-paper scores are inherently paired by
document: each paper is evaluated under both the baseline format and the
candidate combination, and exploiting this within-paper pairing yields
substantially greater statistical power than treating the two score sets as
independent samples, an advantage that is meaningful at $n = 200$.
Beyond these distributional considerations, the Wilcoxon signed-rank test
assesses the sign and relative magnitude of per-paper differences rather than
their raw arithmetic values, preventing individual papers with unusually large
performance gaps from disproportionately driving significance decisions.

We control the family-wise error rate (FWER) using the Holm--Bonferroni
step-down correction across all 62 pairwise comparisons per metric.
Unlike the more conservative Bonferroni correction, which applies a uniform
penalty of $\alpha / m$ to every one of the $m$ comparisons simultaneously,
Holm--Bonferroni ranks all $p$-values from smallest to largest and applies
sequentially decreasing corrections, halting once a comparison fails to reach
the adjusted threshold.
This step-down approach maintains strict FWER control at $\alpha = 0.05$ while
preserving markedly greater statistical power for the most significant
comparisons, a meaningful difference when evaluating 62 format combinations
against a single baseline.
Holm-corrected $p$-values and effect sizes (matched-pairs rank-biserial
correlation $r$) for all comparisons are reported in Table~\ref{tab:combinations}
and Appendix~\ref{app:all_format_combinations}; results are summarized using
significance markers ($^{*}p < 0.05$, $^{**}p < 0.01$, $^{***}p < 0.001$,
\textit{ns} = not significant).

\section{Comprehensive Format-Combination Evaluation Results}
\label{app:all_format_combinations}

This appendix provides the exhaustive evaluation results for all 63 possible format combinations utilized in our URL extraction pipeline, expanding upon the representative subsets highlighted in the main text. 

\subsection{Evaluation Parameters and Metrics}
Performance is benchmarked against the full ground-truth corpus consisting of $2,338$ valid total URLs ($\#\text{V}$) and a specialized subset of $488$ Open Access Data and Software (OADS) URLs ($\#\text{VO}$). For every format combination, we report three core performance metrics: Precision ($P$), Recall ($R$), and the $F_1$-Score ($F_1$). To ensure statistical rigor and account for document-level variances, variability is reported via bootstrap 95\% confidence intervals [95\% CI] enclosed in brackets. These intervals are computed using paired document-level resampling over $B=10{,}000$ iterations.

\begin{figure*}[ht]
  \centering
  \setlength{\fboxsep}{0.5pt}\setlength{\fboxrule}{0.0pt}
  \fbox{\includegraphics[trim=1 1 1 1, clip,
    width=1.0\linewidth]{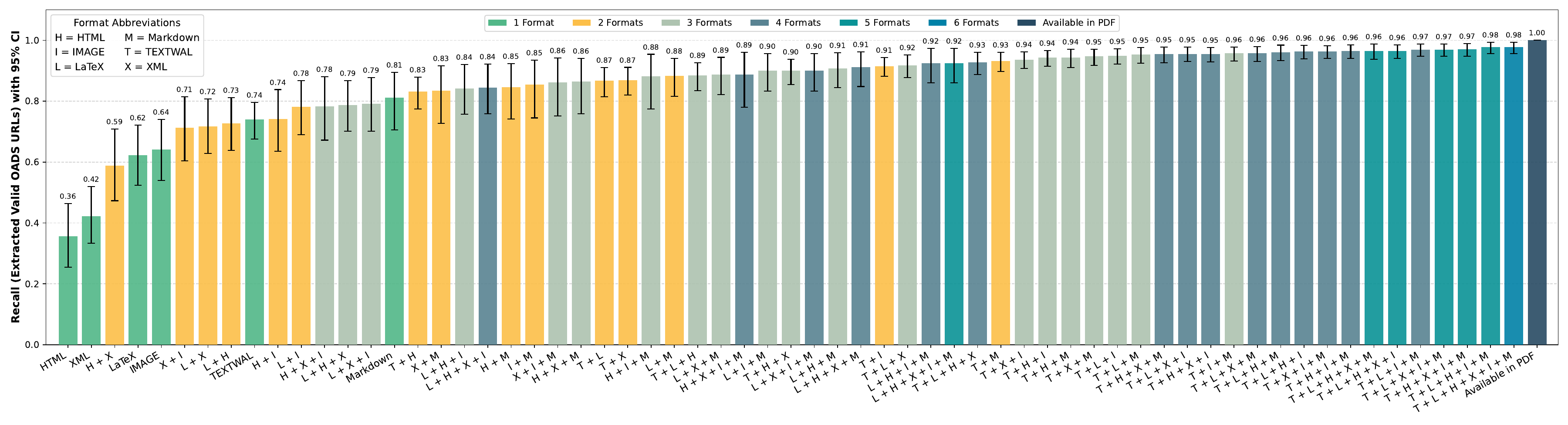}}
  \caption{URL extraction performance across all 63 format combinations for all valid \textbf{OADS URLs}, ordered by recall. The X-axis represents the format combinations, where single-format scenarios are labeled by their full name (e.g., XML, HTML, IMAGE for Qwen2-VL), and multi-format combinations are abbreviated. The "Available in PDF" category serves as the ground truth. Error bars represent 95\% confidence intervals.
  }
  \label{fig:oads_recall_63}
\end{figure*}

\begin{table*}[htbp]
\centering
\caption{Characteristic failure modes by format. Values are reported as \emph{misses/ground truth} (percentage). Ground-truth counts are: References (1781), Title \& Authors (185), Methodology (171), Results \& Discussion (72), Introduction (40), Related Work (24), Header (21), Abstract (14), Footer (12); embedded references (1727), embedded footnotes (275), split across lines (419), embedded captions (12), split across pages (8); hidden annotations (1295), metadata (79); scheme URLs (2285), path/query URLs (2080).}
\label{tab:error_summary_1}
\small
\begin{tabular}{p{0.08\linewidth}p{0.3\linewidth}p{0.25\linewidth}p{0.24\linewidth}}
\toprule
\textbf{Format} & \textbf{Spatial \& Representational Failure} & \textbf{Lexicographical Highlights} & \textbf{Regional Distribution} \\
\midrule

TEXTWAL
& Embedded references: 241/1727 (14.0\%); embedded footnotes: 116/275 (42.2\%); hidden annotations: 168/1295 (13.0\%).
& Scheme URLs: 433/2285 (18.9\%); path/query URLs: 278/2080 (13.4\%).
& References: 252/1781 (14.1\%); Title \& Authors: 81/185 (43.8\%); Methodology: 39/171 (22.8\%). \\

\addlinespace

HTML
& Hidden annotations: 817/1295 (63.1\%); embedded references: 1052/1727 (60.9\%); split lines: 231/419 (55.1\%).
& Scheme URLs: 1466/2285 (64.2\%); path/query URLs: 1351/2080 (65.0\%).
& References: 1077/1781 (60.5\%); Title \& Authors: 169/185 (91.4\%); Methodology: 121/171 (70.8\%). \\

\addlinespace

XML
& Hidden annotations: 1266/1295 (97.8\%); embedded references: 1152/1727 (66.7\%); embedded footnotes: 251/275 (91.3\%).
& Scheme URLs: 1620/2285 (70.9\%); path/query URLs: 1475/2080 (70.9\%).
& References: 1188/1781 (66.7\%); Title \& Authors: 185/185 (100\%); Methodology: 141/171 (82.5\%). \\

\addlinespace

LaTeX
& Hidden annotations: 868/1295 (67.0\%); embedded references: 785/1727 (45.5\%); embedded footnotes: 109/275 (39.6\%).
& Scheme URLs: 1131/2285 (49.5\%); path/query URLs: 1010/2080 (48.6\%).
& References: 818/1781 (45.9\%); Title \& Authors: 181/185 (97.8\%); Methodology: 51/171 (29.8\%). \\

\addlinespace

Markdown
& Hidden annotations: 718/1295 (55.4\%); embedded references: 583/1727 (33.8\%); embedded footnotes: 77/275 (28.0\%).
& Scheme URLs: 843/2285 (36.9\%); path/query URLs: 738/2080 (35.5\%).
& References: 600/1781 (33.7\%); Title \& Authors: 173/185 (93.5\%); Methodology: 24/171 (14.0\%). \\

\addlinespace

VLM (Qwen)
& Hidden annotations: 1250/1295 (96.5\%); embedded references: 1224/1727 (70.9\%); split lines: 223/419 (53.2\%).
& Scheme URLs: 1507/2285 (66.0\%); path/query URLs: 1405/2080 (67.5\%).
& References: 1257/1781 (70.6\%); Title \& Authors: 181/185 (97.8\%); Methodology: 48/171 (28.1\%). \\

\bottomrule
\end{tabular}
\end{table*}

\subsection{Reading the Table}
The comprehensive results are organized by format grouping, with the highest-performing configuration within each operational group highlighted in \textbf{bold}. For comparative baselines, combinations mirroring previous-study configurations (e.g., LaTeX + HTML + XML combinations) are designated with a dagger symbol ($\dag$).

Statistical significance testing is performed against the highest-performing single-format baseline using a two-sided Wilcoxon signed-rank test. To counteract the risk of Type I errors resulting from multiple pairwise evaluations, $p$-values are adjusted using the Holm--Bonferroni correction framework, annotated as follows:
\begin{itemize}[leftmargin=*]
    \item $^{***} p < 0.001$ (Highly significant performance variance)
    \item $^{**} p < 0.01$ (Statistically significant)
    \item $^{*} p < 0.05$ (Marginally significant)
    \item \textit{ns} = not significant ($p \ge 0.05$)
\end{itemize}


\renewcommand{\arraystretch}{1.1}
\begin{table*}[ht]
\myfootnotesize
\centering
\caption{URL extraction performance across all 63 format combinations.
  \#V = valid URLs extracted;
  P [95\% CI] = Precision with bootstrap 95\% confidence interval;
  R [95\% CI] = Recall with bootstrap 95\% confidence interval
  ($B=10{,}000$, paired document-level resampling);
  F1 [95\% CI] = F1-Score with bootstrap 95\% confidence interval.
  Best F1 and OADS Recall per group in \textbf{bold};
  Sig.\ vs.\ best single-format (Wilcoxon, Holm--Bonferroni corrected):
  $^{***}p<0.001$, $^{**}p<0.01$, $^{*}p<0.05$, \textit{ns}.}
\label{tab:combinations_full}
\begin{tabular}{p{0.15\linewidth}llllrr|llrr}
\toprule
\multirow{2}{=}{\textbf{Format Combination}}
  & \multicolumn{6}{c|}{\textbf{Valid URLs (All)}}
  & \multicolumn{4}{c}{\textbf{OADS URLs}} \\
\cmidrule(lr){2-7} \cmidrule(l){8-11}
  & \textbf{\#V}
  & \textbf{P [95\% CI]}
  & \textbf{R [95\% CI]}
  & \textbf{F1 [95\% CI]}
  & \textbf{$p$-value}
  & \textbf{Sig.}
  & \textbf{\#V}
  & \textbf{R [95\% CI]}
  & \textbf{$p$-value}
  & \textbf{Sig.} \\
\midrule
  \quad \textbf{TEXTWAL} & 1,887 & 0.60 [0.51,0.66] & 0.81 [0.75,0.85] & \textbf{0.69 [0.61,0.74]} & - & \textit{ns} & 361 & 0.74 [0.68,0.80] & $<0.001$ & $^{***}$ \\
  \quad Markdown & 1,474 & 0.69 [0.64,0.73] & 0.63 [0.53,0.73] & 0.66 [0.60,0.71] & 0.309 & \textit{ns} & 396 & \textbf{0.81 [0.71,0.89]} & - & \textit{ns} \\
  \quad LaTeX & 1,174 & 0.74 [0.67,0.80] & 0.50 [0.38,0.63] & 0.60 [0.49,0.69] & $<0.001$ & $^{***}$ & 304 & 0.62 [0.52,0.72] & $<0.001$ & $^{***}$ \\
  \quad HTML & 827 & 0.49 [0.33,0.64] & 0.35 [0.23,0.49] & 0.41 [0.27,0.55] & $<0.001$ & $^{***}$ & 174 & 0.36 [0.26,0.46] & $<0.001$ & $^{***}$ \\
  \quad IMAGE & 789 & 0.11 [0.08,0.15] & 0.34 [0.26,0.44] & 0.17 [0.13,0.21] & $<0.001$ & $^{***}$ & 313 & 0.64 [0.54,0.74] & 0.010 & $^{**}$ \\
  \quad XML & 670 & 0.69 [0.63,0.74] & 0.29 [0.22,0.38] & 0.41 [0.33,0.50] & $<0.001$ & $^{***}$ & 206 & 0.42 [0.33,0.52] & $<0.001$ & $^{***}$ \\
\midrule
  \quad T + M & 2,098 & 0.55 [0.48,0.62] & 0.90 [0.86,0.92] & 0.68 [0.62,0.74] & $<0.001$ & $^{***}$ & 455 & \textbf{0.93 [0.90,0.96]} & 0.005 & $^{**}$ \\
  \quad \textbf{T + L} & 2,059 & 0.57 [0.50,0.63] & 0.88 [0.84,0.91] & \textbf{0.69 [0.63,0.75]} & $<0.001$ & $^{***}$ & 423 & 0.87 [0.81,0.91] & 1.000 & \textit{ns} \\
  \quad T + I & 2,059 & 0.21 [0.16,0.27] & 0.88 [0.85,0.91] & 0.34 [0.27,0.42] & $<0.001$ & $^{***}$ & 446 & 0.91 [0.88,0.94] & 0.330 & \textit{ns} \\
  \quad T + H & 2,049 & 0.49 [0.42,0.55] & 0.88 [0.84,0.91] & 0.63 [0.57,0.68] & $<0.001$ & $^{***}$ & 406 & 0.83 [0.77,0.88] & 0.025 & $^{*}$ \\
  \quad \textbf{T + X} & 2,035 & 0.57 [0.49,0.63] & 0.87 [0.83,0.90] & \textbf{0.69 [0.62,0.74]} & $<0.001$ & $^{***}$ & 424 & 0.87 [0.82,0.91] & 1.000 & \textit{ns} \\
  \quad H + M & 1,767 & 0.54 [0.47,0.61] & 0.76 [0.66,0.84] & 0.63 [0.56,0.69] & 1.000 & \textit{ns} & 413 & 0.85 [0.74,0.92] & 0.126 & \textit{ns} \\
  \quad L + M & 1,741 & 0.66 [0.60,0.70] & 0.74 [0.67,0.82] & 0.70 [0.65,0.74] & 1.000 & \textit{ns} & 431 & 0.88 [0.82,0.94] & 0.007 & $^{**}$ \\
  \quad I + M & 1,541 & 0.18 [0.14,0.23] & 0.66 [0.56,0.77] & 0.28 [0.23,0.35] & 1.000 & \textit{ns} & 417 & 0.85 [0.74,0.94] & 0.005 & $^{**}$ \\
  \quad X + M & 1,523 & 0.62 [0.57,0.66] & 0.65 [0.55,0.76] & 0.63 [0.58,0.69] & 1.000 & \textit{ns} & 407 & 0.83 [0.73,0.92] & 0.126 & \textit{ns} \\
  \quad L + H & 1,471 & 0.54 [0.45,0.62] & 0.63 [0.51,0.74] & 0.58 [0.48,0.67] & $<0.001$ & $^{***}$ & 355 & 0.73 [0.64,0.81] & 0.003 & $^{**}$ \\
  \quad L + I & 1,369 & 0.17 [0.12,0.23] & 0.59 [0.47,0.70] & 0.26 [0.20,0.33] & 0.071 & \textit{ns} & 381 & 0.78 [0.69,0.87] & 1.000 & \textit{ns} \\
  \quad H + I & 1,348 & 0.16 [0.11,0.21] & 0.58 [0.45,0.71] & 0.25 [0.19,0.32] & 0.075 & \textit{ns} & 362 & 0.74 [0.64,0.84] & 0.983 & \textit{ns} \\
  \quad L + X & 1,312 & 0.66 [0.58,0.72] & 0.56 [0.44,0.68] & 0.61 [0.51,0.69] & $<0.001$ & $^{***}$ & 350 & 0.72 [0.63,0.81] & 0.069 & \textit{ns} \\
  \quad H + X & 1,244 & 0.52 [0.41,0.62] & 0.53 [0.40,0.67] & 0.52 [0.41,0.64] & $<0.001$ & $^{***}$ & 287 & 0.59 [0.47,0.71] & $<0.001$ & $^{***}$ \\
  \quad X + I & 955 & 0.13 [0.10,0.16] & 0.41 [0.32,0.53] & 0.19 [0.15,0.24] & 0.008 & $^{**}$ & 348 & 0.71 [0.60,0.81] & 0.974 & \textit{ns} \\
\midrule
  \quad T + H + M & 2,174 & 0.46 [0.40,0.52] & 0.93 [0.90,0.95] & 0.62 [0.56,0.67] & $<0.001$ & $^{***}$ & 460 & 0.94 [0.91,0.97] & 0.003 & $^{**}$ \\
  \quad T + H + I & 2,157 & 0.20 [0.16,0.25] & 0.92 [0.89,0.94] & 0.33 [0.27,0.40] & $<0.001$ & $^{***}$ & 460 & 0.94 [0.92,0.97] & 0.069 & \textit{ns} \\
  \quad T + L + M & 2,143 & 0.53 [0.46,0.59] & 0.92 [0.89,0.94] & 0.67 [0.61,0.72] & $<0.001$ & $^{***}$ & 465 & 0.95 [0.93,0.98] & $<0.001$ & $^{***}$ \\
  \quad T + H + X & 2,132 & 0.47 [0.41,0.53] & 0.91 [0.88,0.94] & 0.62 [0.56,0.67] & $<0.001$ & $^{***}$ & 439 & 0.90 [0.85,0.94] & 1.000 & \textit{ns} \\
  \quad T + I + M & 2,130 & 0.21 [0.16,0.27] & 0.91 [0.88,0.93] & 0.34 [0.27,0.41] & $<0.001$ & $^{***}$ & 467 & \textbf{0.96 [0.93,0.98]} & $<0.001$ & $^{***}$ \\
  \quad T + X + M & 2,129 & 0.52 [0.45,0.58] & 0.91 [0.88,0.93] & 0.66 [0.60,0.72] & $<0.001$ & $^{***}$ & 462 & 0.95 [0.92,0.97] & 0.001 & $^{**}$ \\
  \quad T + L + I & 2,128 & 0.21 [0.16,0.27] & 0.91 [0.88,0.94] & 0.35 [0.27,0.42] & $<0.001$ & $^{***}$ & 463 & 0.95 [0.92,0.97] & 0.002 & $^{**}$ \\
  \quad T + L + H & 2,125 & 0.47 [0.41,0.53] & 0.91 [0.87,0.94] & 0.62 [0.56,0.67] & $<0.001$ & $^{***}$ & 432 & 0.89 [0.83,0.93] & 1.000 & \textit{ns} \\
  \quad \textbf{T + L + X} & 2,107 & 0.54 [0.47,0.60] & 0.90 [0.86,0.93] & \textbf{0.68 [0.61,0.73]} & $<0.001$ & $^{***}$ & 448 & 0.92 [0.88,0.95] & 0.837 & \textit{ns} \\
  \quad T + X + I & 2,099 & 0.21 [0.16,0.27] & 0.90 [0.87,0.92] & 0.34 [0.27,0.41] & $<0.001$ & $^{***}$ & 457 & 0.94 [0.91,0.96] & 0.015 & $^{*}$ \\
  \quad L + H + M & 1,867 & 0.52 [0.45,0.58] & 0.80 [0.72,0.86] & 0.63 [0.57,0.68] & 1.000 & \textit{ns} & 443 & 0.91 [0.84,0.96] & 0.003 & $^{**}$ \\
  \quad H + I + M & 1,827 & 0.19 [0.14,0.24] & 0.78 [0.68,0.86] & 0.30 [0.24,0.37] & 1.000 & \textit{ns} & 430 & 0.88 [0.77,0.95] & 0.001 & $^{**}$ \\
  \quad H + X + M & 1,809 & 0.50 [0.44,0.57] & 0.77 [0.67,0.86] & 0.61 [0.54,0.67] & 1.000 & \textit{ns} & 422 & 0.86 [0.76,0.94] & 0.010 & $^{**}$ \\
  \quad L + I + M & 1,779 & 0.20 [0.15,0.25] & 0.76 [0.69,0.83] & 0.31 [0.25,0.38] & 1.000 & \textit{ns} & 439 & 0.90 [0.83,0.96] & 0.001 & $^{**}$ \\
  \quad L + X + M & 1,757 & 0.60 [0.54,0.65] & 0.75 [0.68,0.82] & 0.67 [0.61,0.71] & 1.000 & \textit{ns} & 433 & 0.89 [0.82,0.94] & 0.005 & $^{**}$ \\
  \quad L + H + I & 1,626 & 0.18 [0.13,0.23] & 0.70 [0.58,0.81] & 0.28 [0.22,0.35] & 1.000 & \textit{ns} & 411 & 0.84 [0.76,0.92] & 1.000 & \textit{ns} \\
  \quad L + H + X & 1,576 & 0.51 [0.43,0.58] & 0.67 [0.55,0.78] & 0.58 [0.49,0.66] & 0.572 & \textit{ns} & 384 & 0.79 [0.70,0.87] & 1.000 & \textit{ns} \\
  \quad X + I + M & 1,561 & 0.18 [0.14,0.23] & 0.67 [0.56,0.78] & 0.28 [0.22,0.34] & 1.000 & \textit{ns} & 420 & 0.86 [0.75,0.94] & 0.004 & $^{**}$ \\
  \quad H + X + I & 1,446 & 0.16 [0.12,0.21] & 0.62 [0.49,0.75] & 0.26 [0.20,0.32] & 1.000 & \textit{ns} & 382 & 0.78 [0.67,0.88] & 1.000 & \textit{ns} \\
  \quad L + X + I & 1,399 & 0.17 [0.12,0.22] & 0.60 [0.48,0.72] & 0.26 [0.20,0.33] & 0.514 & \textit{ns} & 386 & 0.79 [0.70,0.88] & 1.000 & \textit{ns} \\
\midrule
  \quad T + H + I + M & 2,204 & 0.20 [0.15,0.25] & 0.94 [0.92,0.96] & 0.33 [0.26,0.39] & $<0.001$ & $^{***}$ & 471 & 0.96 [0.94,0.98] & $<0.001$ & $^{***}$ \\
  \quad T + L + H + M & 2,202 & 0.44 [0.39,0.50] & 0.94 [0.92,0.96] & 0.60 [0.55,0.65] & $<0.001$ & $^{***}$ & 469 & 0.96 [0.93,0.98] & $<0.001$ & $^{***}$ \\
  \quad T + H + X + M & 2,202 & 0.44 [0.38,0.49] & 0.94 [0.92,0.96] & 0.60 [0.54,0.65] & $<0.001$ & $^{***}$ & 466 & 0.95 [0.93,0.98] & $<0.001$ & $^{***}$ \\
  \quad T + L + H + I & 2,191 & 0.20 [0.16,0.25] & 0.94 [0.91,0.96] & 0.33 [0.27,0.40] & $<0.001$ & $^{***}$ & 470 & 0.96 [0.94,0.98] & 0.001 & $^{**}$ \\
  \quad T + H + X + I & 2,184 & 0.20 [0.15,0.25] & 0.93 [0.91,0.95] & 0.33 [0.26,0.39] & $<0.001$ & $^{***}$ & 466 & 0.95 [0.93,0.98] & 0.008 & $^{**}$ \\
  \quad T + L + H + X & 2,169 & 0.45 [0.39,0.51] & 0.93 [0.90,0.95] & 0.61 [0.55,0.66] & $<0.001$ & $^{***}$ & 453 & 0.93 [0.89,0.96] & 0.692 & \textit{ns} \\
  \quad T + L + I + M & 2,163 & 0.21 [0.16,0.26] & 0.93 [0.90,0.95] & 0.34 [0.27,0.41] & $<0.001$ & $^{***}$ & 473 & \textbf{0.97 [0.95,0.99]} & $<0.001$ & $^{***}$ \\
  \quad \textbf{T + L + X + M} & 2,154 & 0.50 [0.43,0.56] & 0.92 [0.89,0.94] & \textbf{0.65 [0.58,0.70]} & $<0.001$ & $^{***}$ & 467 & 0.96 [0.93,0.98] & $<0.001$ & $^{***}$ \\
  \quad T + X + I + M & 2,143 & 0.21 [0.16,0.26] & 0.92 [0.89,0.94] & 0.34 [0.27,0.41] & $<0.001$ & $^{***}$ & 470 & 0.96 [0.94,0.98] & $<0.001$ & $^{***}$ \\
  \quad T + L + X + I & 2,140 & 0.21 [0.16,0.26] & 0.92 [0.88,0.94] & 0.34 [0.27,0.41] & $<0.001$ & $^{***}$ & 466 & 0.95 [0.93,0.98] & $<0.001$ & $^{***}$ \\
  \quad L + H + I + M & 1,905 & 0.19 [0.15,0.24] & 0.81 [0.74,0.88] & 0.31 [0.25,0.37] & 0.760 & \textit{ns} & 451 & 0.92 [0.86,0.97] & $<0.001$ & $^{***}$ \\
  \quad L + H + X + M & 1,883 & 0.48 [0.42,0.54] & 0.81 [0.73,0.87] & 0.60 [0.55,0.66] & 1.000 & \textit{ns} & 445 & 0.91 [0.85,0.96] & 0.002 & $^{**}$ \\
  \quad H + X + I + M & 1,844 & 0.19 [0.14,0.23] & 0.79 [0.69,0.87] & 0.30 [0.24,0.36] & 0.670 & \textit{ns} & 433 & 0.89 [0.78,0.96] & $<0.001$ & $^{***}$ \\
  \quad L + X + I + M & 1,784 & 0.19 [0.15,0.24] & 0.76 [0.69,0.84] & 0.31 [0.24,0.37] & 1.000 & \textit{ns} & 439 & 0.90 [0.83,0.96] & 0.001 & $^{**}$ \\
  \quad L + H + X + I & 1,649 & 0.17 [0.13,0.22] & 0.71 [0.59,0.81] & 0.28 [0.22,0.35] & 1.000 & \textit{ns} & 412 & 0.84 [0.76,0.92] & 1.000 & \textit{ns} \\
\midrule
  \quad T + L + H + I + M & 2,222 & 0.20 [0.15,0.24] & 0.95 [0.93,0.97] & 0.33 [0.26,0.39] & $<0.001$ & $^{***}$ & 477 & \textbf{0.98 [0.96,0.99]} & $<0.001$ & $^{***}$ \\
  \quad T + H + X + I + M & 2,216 & 0.20 [0.15,0.24] & 0.95 [0.93,0.96] & 0.32 [0.26,0.39] & $<0.001$ & $^{***}$ & 474 & 0.97 [0.95,0.99] & $<0.001$ & $^{***}$ \\
  \quad \textbf{T + L + H + X + M} & 2,213 & 0.42 [0.37,0.47] & 0.95 [0.93,0.96] & \textbf{0.58 [0.53,0.63]} & $<0.001$ & $^{***}$ & 471 & 0.96 [0.94,0.99] & $<0.001$ & $^{***}$ \\
  \quad T + L + H + X + I & 2,201 & 0.20 [0.15,0.24] & 0.94 [0.92,0.96] & 0.33 [0.26,0.39] & $<0.001$ & $^{***}$ & 471 & 0.96 [0.94,0.98] & $<0.001$ & $^{***}$ \\
  \quad T + L + X + I + M & 2,165 & 0.20 [0.15,0.26] & 0.93 [0.90,0.95] & 0.33 [0.26,0.40] & $<0.001$ & $^{***}$ & 473 & 0.97 [0.95,0.99] & $<0.001$ & $^{***}$ \\
  \quad L + H + X + I + M & 1,910 & 0.19 [0.14,0.23] & 0.82 [0.74,0.88] & 0.30 [0.24,0.37] & 0.514 & \textit{ns} & 451 & 0.92 [0.86,0.97] & $<0.001$ & $^{***}$ \\
\midrule
  \quad \textbf{T + L + H + X + I + M} & 2,224 & 0.19 [0.15,0.24] & 0.95 [0.93,0.97] & \textbf{0.32 [0.26,0.38]} & $<0.001$ & $^{***}$ & 477 & \textbf{0.98 [0.96,0.99]} & $<0.001$ & $^{***}$ \\
\bottomrule
\end{tabular}
\end{table*}

\section{Error Analysis: Format-Specific Failure Modes}
\label{sec:discussion:error:summary}

Table~\ref{tab:error_summary_1} consolidates the characteristic failure modes across formats. The most practically consequential finding is the \emph{complementarity} of failure modes: plain text and TEXTWAL fail predominantly on spatial and representational complexity; XML on document location (body text vs.\ references); LaTeX on macro convention; Markdown on layout edge cases; VLMs on content they cannot visually resolve.
This complementarity is precisely the mechanism by which multi-format combinations (Section~\ref{sec:results:combinations}) achieve recall substantially above any single format.
Practitioners should audit their target corpus for the failure modes most likely to be present---complex layouts, annotation-heavy workflows, or scanned content---and select format combinations to specifically address those failure modes.

\end{document}